\documentclass[preprint,review,11pt]{elsarticle}
\usepackage[T1]{fontenc}
\usepackage{array}
\usepackage{textcomp}
\usepackage{multirow}
\usepackage{amsmath}
\usepackage{graphicx}
\usepackage{kantlipsum}
\usepackage{esint}
\usepackage{cuted}
\usepackage{stackrel}
\usepackage{graphicx}
\usepackage{mathrsfs}
\usepackage{amstext}
\usepackage{lineno,hyperref}
\modulolinenumbers[5]

\journal{Mechanical Systems and Signal Processing}

\makeatletter

\providecommand{\tabularnewline}{\\}

\@ifundefined{showcaptionsetup}{}{%
	\PassOptionsToPackage{caption=false}{subfig}}
\usepackage{subfig}
\makeatother

\usepackage{babel}
\begin{document}
	
	\begin{frontmatter}{}
		
		\title{Empirical Fourier Decomposition: An Accurate Adaptive Signal Decomposition
			Method }
		
		\author[WUT,UC]{Wei~Zhou}
		\ead{weizhou@mail.uc.edu}
		\author[WUT,SYU]{Zhongren~Feng}
		\ead{ fengz515@163.com }
		\author[UC]{Y.F.~Xu\corref{cor1}}
		\ead{xu2yf@uc.edu}
		\author[WUT]{Xiongjiang~Wang}
		\ead{wangxiongjiang@whut.edu.cn}
		\author[WUT]{Hao~Lv}
		\ead{liberty.hunter@whut.edu.cn}

		\address[WUT]{School of Civil Engineering and Architecture, Wuhan University
			of Technology, Wuhan 430070, P.R. China}
		
		\address[UC]{Department of Mechanical and Materials Engineering, University of Cincinnati, Cincinnati, OH 45221, USA}
		
		\address[SYU]{School of Urban Construction, Wuchang Shouyi University, Wuhan
			430064, P.R. China}
		
		\cortext[cor1]{Corresponding author. Tel:~+1-513-556-2755.}
		\begin{abstract}
			Signal decomposition is an effective tool to assist the identification
			of modal information in time-domain signals. Two signal decomposition
			methods, including the empirical wavelet transform (EWT) and Fourier
			decomposition method (FDM), have been developed based on Fourier theory.
			However, the EWT can suffer from a mode mixing problem for signals
			with closely-spaced modes and decomposition results by FDM can suffer
			from an inconsistency problem. An accurate adaptive signal decomposition
			method, called the empirical Fourier decomposition (EFD), is proposed
			to solve the problems in this work. The proposed EFD combines the
			uses of an improved frequency spectrum segmentation technique and
			an ideal filter bank. The segmentation technique can solve the inconsistency
			problem by predefining the number of modes in a signal to be decomposed
			and filter functions in the ideal filter bank have no transition phases,
			which can solve the mode mixing problem. Numerical investigations
			are conducted to study the accuracy of the EFD. It is shown that the
			EFD can yield accurate and consistent decomposition results for signals
			with multiple non-stationary modes and those with closely-spaced modes,
			compared with decomposition results by the EWT, FDM, variational mode
			decomposition and empirical mode decomposition. It is also shown that
			the EFD can yield accurate time-frequency representation results and
			it has the highest computational efficiency among the compared methods. 
		\end{abstract}
		
	\end{frontmatter}{}
	
	\linenumbers

\section{Introduction\label{sec:1}}

Signal decomposition is a widely used numerical tool in different
fields, such as biomedical signal analysis \citep{bhattacharyya2017multivariate},
seismic signal analysis \citep{li2018seismic}, mechanical vibration
signal analysis \citep{pan2016mono,chen2019adaptive}, and speech
enhancement \citep{upadhyay2017speech}. Time-domain signals that
derive from a physical system usually comprise several superposed
components, which are referred to as modes \citep{iatsenko2015nonlinear},
and the modes can encompass meaningful frequency-domain information
of the signals, i.e. modal information. Hence, it is crucial to obtain
signal decomposition results with high accuracy and efficiency.

In the past few decades, several signal decomposition methods have
been developed, and the empirical mode decomposition (EMD)\citep{huang1998empirical}
is one of the most significant methods, even though its mathematical
understanding is limited and it has some known shortcomings, such
as robustness of mode mixing \citep{rilling2003empirical} and end
effects \citep{rato2008hht}. Improved versions of the EMD have been
developed to overcome the shortcomings. The ensemble EMD \citep{wu2009ensemble}
has been developed by adding white noise with finite amplitudes to
alleviate the mode mixing and end effects problems. The complete ensemble
EMD \citep{torres2011complete} was developed to further improve the
EMD by adding completeness and a full data-driven number of modes,
which are missing in the ensemble EMD. Li et al. presented a time-varying
filter technique to solve the mode mixing problem \citep{li2017time}.
However, these EMD methods cannot fundamentally solve the mode mixing
and end effects problems. The variational mode decomposition (VMD)
\citep{dragomiretskiy2013variational} is a non-recursive signal decomposition
method, which was developed based on a generalization of Wiener filters.
However, the VMD can fail for non-stationary time-domain signals with
chirp modes that have overlapping frequency ranges. To avoid this
potential failure, McNeill \citep{mcneill2016decomposing} proposed
the use of an optimized objective function with constraints on short-time
narrow-band modes, and Chen et al. \citep{chen2017nonlinear} exploited
a complete variational framework to generalize the VMD. The empirical
wavelet transform (EWT) employs an adaptive wavelet filter bank based
on segments of Fourier spectra \citep{gilles2013empirical}. The workability
of the EWT has been improved in Refs. \citep{luo2018revised,xin2019operational,amezquita2015new}
to eliminate its requirement for a high signal-to-noise ratio in a
signal to be decomposed. However, transition phases between filter
functions in a wavelet filter bank in the EWT can lead to the model
mixing problem for signals with closely-spaced modes. Besides, the
first decomposed mode by the EWT can correspond to a trivial residual,
with which the number of segments is not equal to that of decomposed
modes. Fourier decomposition method (FDM) \citep{singh2017fourier}
is an adaptive non-stationary, non-linear signal decomposition method
that can decompose a zero-mean signal into a set of Fourier intrinsic
band functions (FIBFs) based on Fourier theory and Hilbert transform.
Several limitations of the FDM have been identified. To obtain a FIBF,
two frequency scan techniques were developed. One is called the low-to-high
(LTH) technique and the other is high-to-low (HTL) technique. The
LTH and HTL techniques recursively estimate upper and lower bounds
of FIBFs, respectively. However, decomposition results by the FDM
with the two frequency scan techniques, i.e., FDM-HTL and FDM-LTH,
can be inconsistent and one cannot determine which decomposition results
are correct. Further, the two frequency scan techniques are both iterative
and can require long computation times for FDM. 

In this work, the EWT and the FDM are briefly reviewed: the segmentation
technique and construction of a wavelet filter bank in the EWT are
described, and the construction of FIBFs and the two frequency scan
techniques of the FDM are described. A new adaptive signal decomposition
method, called the empirical Fourier decomposition (EFD), is proposed
to solve the aforementioned problems of the EWT and FDM. The EFD combines
an improved segmentation technique and an ideal filter bank \citep{alan2009discrete}.
The improved segmentation technique has an adaptative sorting process
to yield segmentation results that are less adversely affected by
noise and contains meaningful modal information, and the ideal filter
bank, whose filter functions have no transition phases, facilitates
accurate decomposition for signals with closely-spaced modes. Numerical
investigations are conducted to study the accuracy of decomposition
results by the EFD for two non-stationary signals and two signals
with closely-spaced modes by comparing with decomposition results
by the EMD, VMD, EWT and FDM. In addition, the accuracy of time-frequency
representations (TFRs) and computational efficiency of the EFD are
compared with those associated with the other methods. 

The remnant of the paper is arranged as follows. In Section 2, the
EWT and FDM are briefly reviewed. In Section 3, the proposed EFD is
described. In Section 4, the numerical investigations are presented.
Conclusions and some discussions on future works are presented in
Section 5. 

\section{Reviews of EWT and FDM\label{sec:2}}

\subsection{EWT}

The EWT employs an adaptive wavelet transform algorithm based on segments
of Fourier spectra \citep{gilles2013empirical}. The two most important
steps of the EWT are: (1) use of an adaptive segmentation technique
to divide Fourier spectrum of a signal to be decomposed and (2) construction
of a wavelet filter bank \citep{daubechies1992ten}. Assume that the
spectrum is defined on a normalized frequency range $[-\pi,\pi]$.
The segmentation technique and wavelet filter bank for the spectrum
in the frequency range $[0,\pi]$ are described below, and those for
the spectrum in the frequency range $[-\pi,0]$ can be deduced based
on Hermitian symmetry of Fourier spectrum in the normalized frequency
range $[-\pi,\pi]$. 

One segmentation technique for the EWT is the local maxima technique
\citep{gilles2013empirical}, in which the spectrum in $[0,\pi]$
is divided into $N$ contiguous frequency segments. Each segment is
denoted by $S_{n}=[\omega_{n-1},\omega_{n}]$ with $n\in[1,N]$, $\omega_{0}=0$
and $\omega_{N}=\pi$. To determine values of $\omega_{n}$, the first
$N-1$ largest local maxima of the spectrum magnitude are identified.
The frequencies that uniquely correspond to the identified maxima
are re-indexed in descending order and denoted by $[\Omega_{1},\Omega_{2},\ldots\Omega_{N-1}]$
such that $\Omega_{1}<\Omega_{2}<\dots<\Omega_{N-1}$; in addition,
$\Omega_{0}=0$ is defined. The value of $\omega_{n}$ is expressed
by 
\begin{equation}
\omega_{n}=\frac{\Omega_{n-1}+\Omega_{n}}{2},n\in[1,N-1]\label{eq:eq1}
\end{equation}
which concludes the local maxima technique. As an alternative to the
local maxima technique, the lowest minima technique \citep{gilles20142d}
was developed for the EWT: the spectrum division and frequency reindexing
procedures, which are the same as those in the local maxima technique,
are first carried out. Then the minimum of the spectrum magnitude
in the frequency range $\left[\Omega_{n-1},\Omega_{n}\right]$ is
identified and the value of $\omega_{n}$ is determined by 
\begin{equation}
\omega_{n}=\underset{\omega}{\arg}\min X_{n}(\omega)\label{eq:eq2}
\end{equation}
where $X_{n}(\omega)$ denotes spectrum magnitudes between in $\left[\Omega_{n-1},\Omega_{n}\right]$
and $\arg\min(\cdot)$ denotes argument of the minimum, respectively,
which concludes the lowest minima technique. 

\begin{figure}
\begin{centering}
\includegraphics[scale=0.6]{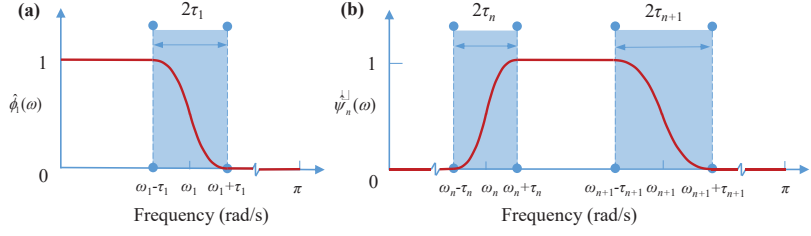}
\par\end{centering}
\caption{\label{fig:1}Graphical illustrations of (a) $\hat{\phi}_{1}(\omega)$
and (b) $\hat{\psi}_{n}(\omega)$. Shaded parts are transition phases.}
\end{figure}

The wavelet filter bank is then constructed, and it consists of an
empirical scaling function $\hat{\phi}_{1}(\omega)$ and a series
of empirical wavelet functions $\hat{\psi}_{n}(\omega)$, which are
expressed by
\begin{equation}
\hat{\phi}_{1}(\omega)=\begin{cases}
1 & \text{if }\arrowvert\omega\arrowvert\leq\omega_{1}-\tau_{1}\\
\cos\left[\frac{\pi}{2}\beta\left(\frac{1}{2\tau_{1}}\left(\tau_{1}+\arrowvert\omega\arrowvert-\omega_{1}\right)\right)\right] & \text{if }\omega_{1}-\tau_{1}\leq\arrowvert\omega\arrowvert\leq\omega_{1}+\tau_{1}\\
0 & \text{otherwise}
\end{cases}\label{eq:eq3}
\end{equation}
and 
\begin{equation}
\psi_{n}(\omega)=\begin{cases}
1 & \text{if }\omega_{n}+\tau_{n}\leq\arrowvert\omega\arrowvert\leq\omega_{n+1}-\tau_{n+1}\\
\cos\left[\frac{\pi}{2}\beta\left(\frac{1}{2\tau_{n+1}}\left(\tau_{n+1}+\arrowvert\omega\arrowvert-\omega_{n+1}\right)\right)\right] & \text{if }\omega_{n+1}-\tau_{n+1}\leq\arrowvert\omega\arrowvert\leq\omega_{n+1}+\tau_{n+1}\\
\sin\left[\frac{\pi}{2}\beta\left(\frac{1}{2\tau_{n}}\left(\tau_{n}+\arrowvert\omega\arrowvert-\omega_{n}\right)\right)\right] & \text{if }\omega_{n}-\tau_{n}\leq\arrowvert\omega\arrowvert\leq\omega_{n}+\tau_{n}\\
0 & \text{otherwise}
\end{cases}\label{eq:eq4}
\end{equation}
in which respectively, $\hat{\cdot}$ denotes Fourier transform of
a function, $\omega$ the circular frequency, $\beta$ an arbitrary
function and $\tau_{n}$ a parameter that determines the size of the
transition phase \citep{gilles2013empirical} associated with the
$n$-th and $(n+1)$-th segments; the transition phase ranges in $[\omega_{n}-\tau_{n},\omega_{n}+\tau_{n}]$.
One of the most used forms of $\beta$ in Eqs. (\ref{eq:eq3}) and
(\ref{eq:eq4}) with a variable $x$ is \citep{daubechies1992ten}
\begin{equation}
\beta(x)=\begin{cases}
0 & \text{if }x\leq0\\
x^{4}(35-84x+70x^{2}-20x^{3}) & \text{if }0<x<1\\
1 & \text{\text{if }}x\geq1
\end{cases}\label{eq:eq5}
\end{equation}
The parameter $\tau_{n}$ is calculated by 
\begin{equation}
\tau_{n}=\gamma\omega_{n}\label{eq:eq6}
\end{equation}
where $\gamma$ is a sufficiently small parameter, so that it can
prevent overlapping between boundaries of non-zero $\hat{\phi}_{1}(\omega)$
and $\hat{\psi}_{n}(\omega)$. A criterion for an acceptable value
of $\gamma$ is:
\begin{equation}
\gamma<\underset{n}{\min}(\frac{\omega_{n+1}-\omega_{n}}{\omega_{n+1}+\omega_{n}})\label{eq:eq7}
\end{equation}
for all $n$ values, and its value can be determined by
\begin{equation}
\gamma=(\frac{R-1}{R})\underset{n}{\min}(\frac{\omega_{n+1}-\omega_{n}}{\omega_{n+1}+\omega_{n}})\label{eq:eq8}
\end{equation}
where $R$ is the number of discrete data in the signal to be decomposed.
The determination of $\hat{\phi}_{1}(\omega)$ and $\hat{\psi}_{n}(\omega)$
concludes the construction of the wavelet filter bank. Graphical illustrations
of $\hat{\phi}_{1}(\omega)$ and $\hat{\psi}_{n}(\omega)$ are shown
in Figs. \ref{fig:1}(a) and (b), respectively.

After applying a segmentation technique and constructing a filter
bank, a decomposed signal can be reconstructed as

\begin{align}
\tilde{f}(t) & =W_{f}^{\varepsilon}(0,t)*\phi_{1}(t)+\sum_{n=1}^{N-1}W_{f}^{\varepsilon}(n,t)*\psi_{n}(t)\label{eq:9}
\end{align}
where the asterisk $*$ denotes the convolution of two functions,
$W_{f}^{\varepsilon}(0,t)$ and $W_{f}^{\varepsilon}(n,t)$ are called
the approximation coefficient function and detail coefficient function,
respectively. The function $W_{f}^{\varepsilon}(0,t)$ is expressed
by

\begin{align}
W_{f}^{\varepsilon}(0,t) & =F^{-1}(\hat{f}(\omega)\hat{\phi}_{1}(\omega))\nonumber \\
 & =\int_{-\omega_{1}-\tau_{1}}^{\omega_{1}+\tau_{1}}f(\tau)\bar{\phi}_{1}(\tau-t)\text{d}\tau
\end{align}
where the overbar denotes complex conjugation and $F^{-1}$ denotes
the inverse Fourier transform of a function. Note that $F^{-1}(\hat{\phi}_{1}(\omega))=\phi_{1}(t)$
and $F^{-1}(\hat{\psi}_{n}(\omega))=\psi_{n}(t)$. The function $W_{f}^{\varepsilon}(n,t)$
is expressed by 

\begin{align}
W_{f}^{\varepsilon}(n,t) & =F^{-1}(\hat{f}(\omega)\hat{\psi}_{n}(\omega))\nonumber \\
 & =\int_{-\omega_{n+1}-\tau_{n+1}}^{\tau_{n}-\omega_{n}}f(\tau)\bar{\psi}_{n}(\tau-t)\text{d}\tau+\int_{\omega_{n}-\tau_{n}}^{\omega_{n+1}+\tau_{n+1}}f(\tau)\bar{\psi}_{n}(\tau-t)\text{d}\tau
\end{align}

Resulting decomposed components of the signal can be expressed by

\begin{equation}
f_{0}(t)=W_{f}^{\varepsilon}(0,t)*\phi_{1}(t)
\end{equation}

and

\begin{equation}
f_{n}(t)=W_{f}^{\varepsilon}(n,t)*\psi_{n}(t)
\end{equation}

A step-by-step description of the EWT for a signal $f(t)$ is provided
as follows 

\textbf{Step 1.} Obtain a Fourier spectrum of $f(t)$ using Fourier
transform. 

\textbf{Step 2.} Segment the spectrum in Step 1 using a segmentation
technique, such as the local maxima technique and lowest minimum technique. 

\textbf{Step 3.} Construct a wavelet filter bank based on the frequency
segments in Step 2. 

\textbf{Step 4.} Express approximation and detail coefficient functions
based on the wavelet filter bank in Step 3. 

\textbf{Step 5.} Decompose $f(t)$ and reconstruct it using in Eq.
(\ref{eq:9}). 

The EWT can yield accurate decomposition results when $f(t)$ does
not have closely-spaced modes. However, when $f(t)$ has closely-spaced
modes, a mode mixing problem can occur due to the transition phase,
and the problem can become more serious when the closely-spaced modes
exist in high frequencies as $\tau_{n}$ in Eq. (\ref{eq:eq6}) becomes
larger for higher frequencies.

\subsection{FDM }

Assume that $f(u)$ is zero-mean, discrete, and a length-limited signal
within one period $U$, which is an even integer; the fundamental
frequency of $f(u)$ can be expressed by 

\begin{equation}
\varphi_{0}=\frac{2\pi}{U}
\end{equation}
In the FDM, $f(u)$ is approximated by a summation of $K$ orthogonal
FIBFs $g_{k}(u)$ \citep{singh2017fourier}:

\begin{equation}
f(u)=\sum_{k=1}^{K}g_{k}(u)\label{eq:15}
\end{equation}
Based on Eq. (\ref{eq:15}), the analytical signal of $f(u)$ can
be expressed by \citep{picinbono1997instantaneous} 

\begin{align}
z(u) & =f(u)+\text{j}H(f(u))\nonumber \\
 & =\sum_{k=1}^{K}\left[g_{k}(u)+\text{j}H(g_{k}(u))\right]\\
 & =\sum_{k=1}^{K}z_{k}(u)\nonumber 
\end{align}
where $H(\cdot)$ denotes Hilbert transform of a function, $\text{j}=\sqrt{-1}$,
and $z_{k}(u)=g_{k}(u)+\text{j}H(g_{k}(u)).$ The term $z_{k}(u)$
can be considered as the analytical signal corresponding to $g_{k}(u)$.
Note that $z(u)$ can be expressed as Fourier series:

\begin{equation}
z(u)=\sum_{m=1}^{U/2-1}a_{m}\text{e}^{\text{j}m\varphi_{0}u}
\end{equation}
where

\begin{equation}
a_{m}=\frac{2}{U}\sum_{u=0}^{U-1}f(u)\text{e}^{-\text{j}m\varphi_{0}u}
\end{equation}
and values of $a_{m}$ can be estimated using discrete Fourier transform.

\begin{figure}
\begin{centering}
\includegraphics[scale=0.6]{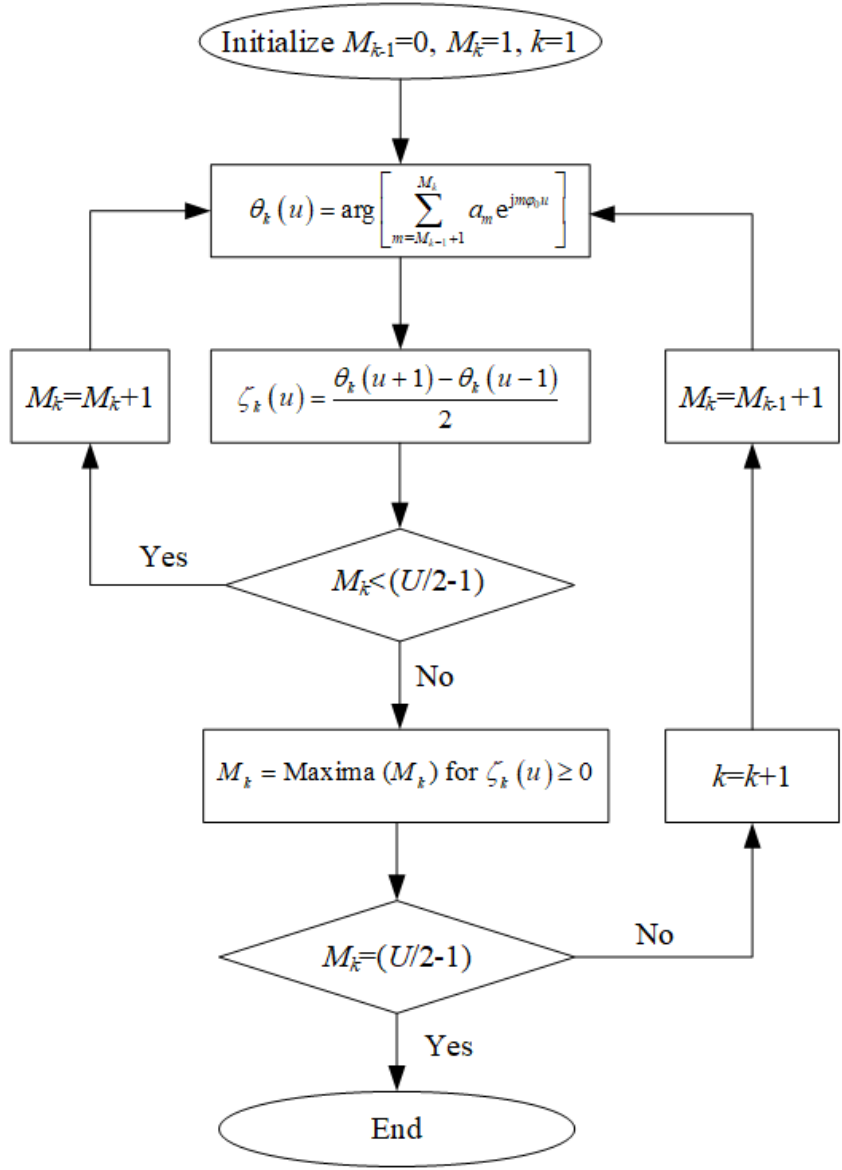}
\par\end{centering}
\caption{\label{fig:2}Flowchart of the LTH technique.}
\end{figure}

\begin{figure}
\begin{centering}
\includegraphics[scale=0.6]{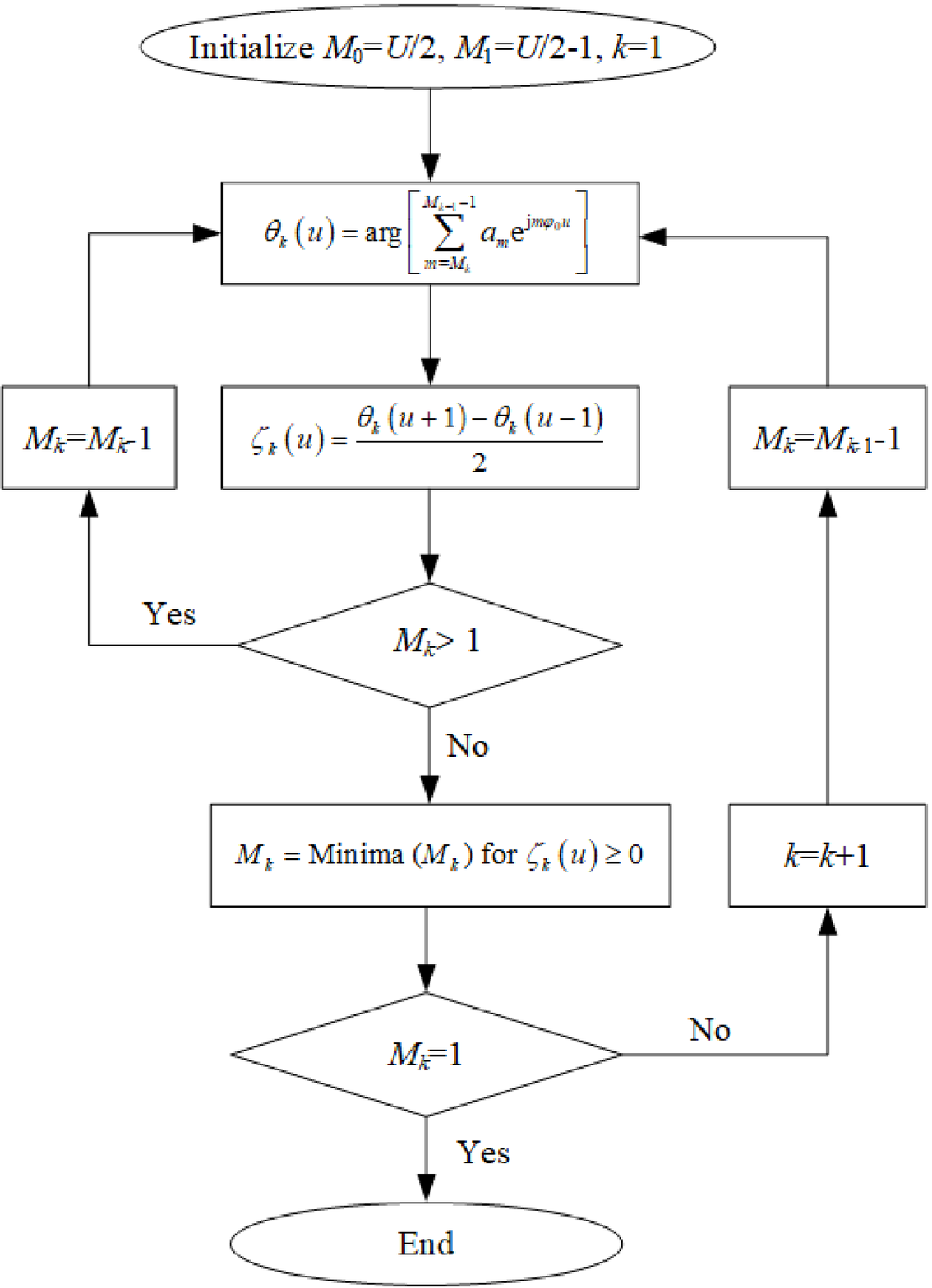}
\par\end{centering}
\caption{\label{fig:3}Flowchart of the HTL technique.}
\end{figure}

The analytical signal $z_{k}(u)$ can be considered as a filtered
signal by Hilbert transform filter \citep{harris2004multirate}, which
is the counterpart of a filter in the wavelet filter bank in the EWT,
and $z_{k}(u)$ can be further expressed by

\begin{equation}
z_{k}(u)=\sum_{m=M_{k-1}+1}^{M_{k}}a_{m}\text{e}^{\text{j}m\varphi_{0}u}\label{eq:19}
\end{equation}
where $M_{k}$ ranges from 1 to $(U/2-1)$ with $M_{0}=0$. Determination
of values of $M_{k}$ is similar to the segmentation in the EWT. In
the FDM, two frequency scan techniques have been proposed to determine
the values of $M_{k}$ in Eq. (\ref{eq:19}), including the LTH technique
and the HTL technique \citep{singh2017fourier}. Flowcharts of the
LTH and HTL techniques are shown in Figs. \ref{fig:2} and \ref{fig:3},
respectively. In the LTH technique, $K$ values of $M_{k}$ are searched
in a forward manner so that $M_{1}<M_{2}<\ldots<M_{k}\ldots<M_{K}$,
with which

\begin{align}
z_{1}(u) & =\sum_{m=M_{0}+1}^{M_{1}}a_{m}\text{e}^{\text{j}m\varphi_{0}u}\nonumber \\
z_{2}(u) & =\sum_{m=M_{1}+1}^{M_{2}}a_{m}\text{e}^{\text{j}m\varphi_{0}u}\\
 & \vdots\nonumber \\
z_{K}(u) & =\sum_{m=M_{K-1}+1}^{M_{K}}a_{m}\text{e}^{\text{j}m\varphi_{0}u}\nonumber 
\end{align}
where $M_{0}=0$ and $M_{K}=U/2-1$. The signal $z_{k}(u)$ can further
be expressed by

\begin{equation}
z_{k}(u)=A_{k}(u)\text{e}^{\text{j}\theta_{k}(u)}
\end{equation}
where

\begin{equation}
A_{k}(u)=\left\Vert \sum_{m=M_{k-1}+1}^{M_{k}}a_{m}\text{e}^{\text{j}m\varphi_{0}u}\right\Vert _{2}
\end{equation}
and

\begin{equation}
\theta_{k}(u)=\arg\left[\sum_{m=M_{k-1}+1}^{M_{k}}a_{m}\text{e}^{\text{j}m\varphi_{0}u}\right]
\end{equation}
denote instantaneous amplitude and phase of $z_{k}(u)$, respectively,
in which $\left\Vert \cdot\right\Vert _{2}$ and $\arg(\cdot)$ calculate
Euclidean norm and argument of a complex quantity, respectively. The
FIBFs $g_{k}(u)$ can be obtained by

\begin{equation}
g_{k}(u)=\text{Re}\left[A_{k}(u)\text{e}^{\text{j}\theta_{k}(u)}\right]
\end{equation}
where $\text{Re}(\cdot)$ is the real part of a function. In the HTL
technique, $K$ values of $M_{k}$ are searched in a backward manner
so that $M_{K}<\ldots M_{k}<M_{k-1}<\ldots<M_{1}$, with which 

\begin{align}
z_{1}(u) & =\sum_{m=M_{1}}^{M_{0}-1}a_{m}\text{e}^{\text{j}m\varphi_{0}u}\nonumber \\
z_{2}(u) & =\sum_{m=M_{2}}^{M_{1}-1}a_{m}\text{e}^{\text{j}m\varphi_{0}u}\\
 & \vdots\nonumber \\
z_{K}(u) & =\sum_{m=M_{K}}^{M_{K-1}-1}a_{m}\text{e}^{\text{j}m\varphi_{0}u}\nonumber 
\end{align}
where $M_{0}=U/2$ and $M_{K}=1$.

A step-by-step description of the FDM for $f(u)$ is as follows 

\textbf{Step 1. }Obtain a Fourier spectrum of $f(u)$ using Fourier
transform. 

\textbf{Step 2. }Express $z(u)$ using the spectrum of $f(u)$ obtained
in Step 1. 

\textbf{Step 3.} Obtain $K$ $z_{k}(u)$ using the LTH or HTL technique. 

\textbf{Step 4. }Obtain FIBFs $g_{k}(u)$ from the real part of $z_{k}(u)$
obtained in Step 3. 

\textbf{Step 5.} Reconstruct $f(u)$ as a summation of FIBFs $g_{k}(u)$
obtained in Step 4.

An issue of the FDM is that its decomposition results using the LTH
technique can be inconsistent with that using the HTL technique, and
the issue will be verified in the numerical investigation in Sec.
4. 

\section{EFD }

Similar to the EWT and FDM, the EFD consists of two critical steps:
an improved segmentation technique and construction of an ideal filter
bank. In the EFD, Fourier spectrum of a signal to be decomposed is
defined on a normalized frequency range $[-\pi,\pi],$ and the improved
segmentation technique and construction of an ideal filter bank for
the spectrum in the frequency range $[0,\pi]$ are described below. 

\subsection{Improved segmentation technique }

The improved segmentation technique is proposed based on the lowest
minima technique \citep{gilles20142d} described in Sec. 2.1. In the
improved segmentation technique, $[0,\pi]$ is divided into $N$ contiguous
frequency segments. Unlike the local maxima and lowest minima techniques,
$\omega_{0}$ and $\omega_{N}$ are not necessarily equal to 0 and
$\pi$, respectively, and their values are determined in an adaptative
sorting process. In the sorting process, Fourier spectrum magnitudes
at $\omega=0$ and $\omega=\pi$ and their local maxima are identified
and extracted to a series. All magnitudes in the series are sorted
in descending order. Frequencies corresponding to the first $N$ largest
values in the sorted series are denoted by $[\Omega_{1},\Omega_{2},\ldots\Omega_{N}]$.
In addition, $\Omega_{0}=0$ and $\Omega_{N+1}=\pi$ are defined.
Boundaries of each segment are determined by 

\begin{align}
\omega_{n} & =\begin{cases}
\underset{\omega}{\arg}\min\check{X}_{n}(\omega) & \text{ if }0\leq n\leq N\text{ and }\Omega_{n}\neq\Omega_{n+1}\\
\Omega_{n} & \text{ if }0\leq n\leq N\text{ and }\Omega_{n}=\Omega_{n+1}
\end{cases}\label{eq:26}
\end{align}
where $\check{X}_{n}(\omega)$ denotes the Fourier spectrum magnitudes
between $\Omega_{n}$ and $\Omega_{n+1}$, which concludes the improved
segmentation technique.

\subsection{Construction of an ideal filter bank }

Both the EWT and FDM consist of a step of constructing a filter bank.
In the EWT, a wavelet filter bank is formed by the empirical scaling
function and wavelet functions. In the FDM, Hilbert transform filter
bank is constructed based on Fourier spectrum of the analytical signal
associated with a signal to be decomposed. In the EFD, an ideal filter
bank is constructed based on frequency segments obtained by the improved
segmentation technique. In each frequency segment, the ideal filter
is a band-pass filter with $\omega_{n-1}$ and $\omega_{n}$ serving
as its cut-off frequencies and it has not transition phases. Hence,
the ideal filter can retain the major Fourier spectrum component in
the segment and all other Fourier spectrum components beyond the segment
are excluded.

\begin{figure}
\begin{centering}
\includegraphics[scale=0.6]{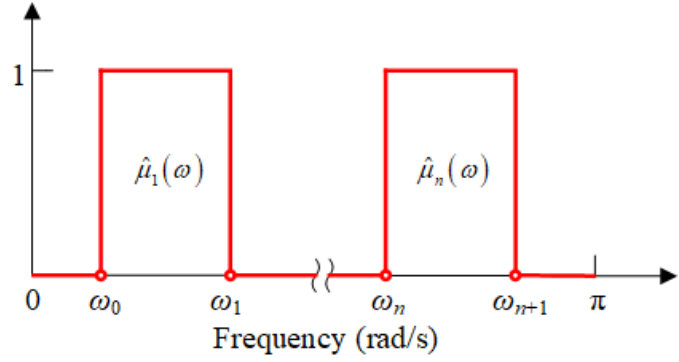}
\par\end{centering}
\caption{\label{fig:4}Graphical illustration of an ideal filter bank of the
EFD.}
\end{figure}

Fourier transform of a signal to be decomposed $f(t)$ is expressed
as 

\begin{equation}
\hat{f}(\omega)=\int_{-\infty}^{\infty}f(t)\text{e}^{-\text{j}\omega t}\text{d}t\label{eq:27}
\end{equation}
An ideal filter bank can be constructed by $\hat{\mu}_{n}(\omega)$
:

\begin{equation}
\hat{\mu}_{n}(\omega)=\begin{cases}
1 & \text{\text{if} \ensuremath{\omega_{n-1}}\ensuremath{\le\mid\omega\mid\le\omega_{n}}}\\
0 & \text{otherwise}
\end{cases}
\end{equation}
where $1\le n\le N$ and values of $\omega_{n}$ are determined by
Eq. (\ref{eq:26}). A graphical illustration of the ideal filter bank
is shown in Fig. \ref{fig:4}. 

Filtered signals that correspond to $\hat{\mu}_{n}(\omega)$ are calculated
by

\begin{equation}
\hat{f_{n}}(\omega)=\hat{\mu}_{n}(\omega)\hat{f}(\omega)=\begin{cases}
\hat{f}(\omega) & \text{\text{if} \ensuremath{\omega_{n-1}}\ensuremath{\le\mid\omega\mid\le\omega_{n}}}\\
0 & \text{otherwise}
\end{cases}
\end{equation}

Decomposed components in the time domain can be obtained using the
inverse Fourier transform:

\begin{align}
f_{n}(t) & =F^{-1}\left[\hat{f_{n}}(\omega)\right]=\int_{-\infty}^{\infty}\hat{f_{n}}(\omega)\text{e}^{\text{j}\omega t}\text{d}\omega\nonumber \\
 & =\int_{-\omega_{n}}^{-\omega_{n-1}}\hat{f}(\omega)\text{e}^{\text{j}\omega t}\text{d}\omega+\int_{\omega_{n-1}}^{\omega_{n}}\hat{f}(\omega)\text{e}^{\text{j}\omega t}\text{d}\omega
\end{align}
The reconstructed signal is calculated as a summation of all decomposed
components:

\begin{equation}
\tilde{f}(t)=\sum_{n=1}^{N}f_{n}(t)
\end{equation}

\begin{figure}
\begin{centering}
\includegraphics[scale=0.6]{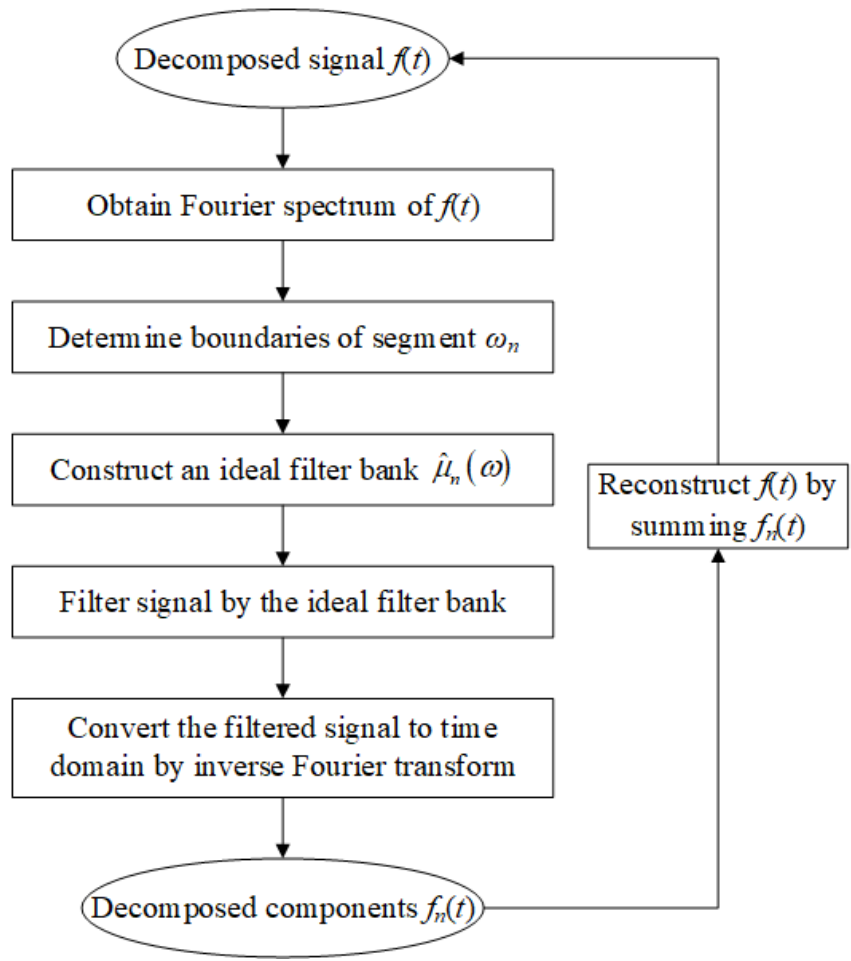}
\par\end{centering}
\caption{\label{fig:5}Flowchart of the EFD.}

\end{figure}

A flowchart of the EFD is shown in Fig. \ref{fig:5} and a step-by-step
description of the EFD is provided as follows. 

\textbf{Step 1.} Obtain Fourier spectrum of a signal to be decomposed
$f(t)$ using Fourier transform. 

\textbf{Step 2.} Determine boundaries of segment $\omega_{n}$ using
the improved segmentation technique based on Fourier spectrum obtained
in Step 1. 

\textbf{Step 3.} Construct an ideal filter bank $\hat{\mu}_{n}(\omega)$
based on $\omega_{n}$ obtained in Step 2. 

\textbf{Step 4.} Obtain filtered signals $\hat{f_{n}}(\omega)$ in
the frequency domain using $\hat{\mu}_{n}(\omega)$ obtained in Step
3. 

\textbf{Step 5.} Obtain decomposed components $f_{n}(t)$ in the time-domain
using inverse Fourier transforms of $\hat{f_{n}}(\omega)$ obtained
in Step 4 

\section{Numerical investigation}

In this section, the effectiveness of the lowest minima and improved
segmentation techniques are compared based on two typical signals
that have multiple modes, denoted by $f_{\text{Sig1}}(t)$ and $f_{\text{Sig2}}(t)$.
Decomposition accuracy of the proposed EFD method is compared with
those of the EWT \citep{gilles2013empirical}, FDM \citep{singh2017fourier},
VMD \citep{dragomiretskiy2013variational} and EMD \citep{huang1998empirical}
methods for two typical non-stationary time-domain signals $f_{\text{Sig3}}(t)$
and $f_{\text{Sig4}}(t)$, and two stationary time-domain signals
$f_{\text{Sig5}}(t)$ and $f_{\text{Sig6}}(t)$, with closely-spaced
modes. For the EFD, EWT and VMD, the numbers of decomposed modes are
listed in Table \ref{tab:1}. 

\begin{table}

\caption{\label{tab:1}Numbers of modes to be decomposed for.}

\centering{}{\footnotesize{}}%
\begin{tabular}{cccc}
\hline 
\multirow{2}{*}{{\footnotesize{}Signal}} & \multicolumn{3}{c}{{\footnotesize{}Decomposition method}}\tabularnewline
\cline{2-4} \cline{3-4} \cline{4-4} 
 & {\footnotesize{}EFD} & {\footnotesize{}EWT} & {\footnotesize{}VMD}\tabularnewline
\hline 
{\footnotesize{}$f_{\text{Sig1}}(t)$} & {\footnotesize{}3} & {\footnotesize{}3} & {\footnotesize{}-}\tabularnewline
{\footnotesize{}$f_{\text{Sig2}}(t)$} & {\footnotesize{}3} & {\footnotesize{}4} & {\footnotesize{}-}\tabularnewline
{\footnotesize{}$f_{\text{Sig3}}(t)$} & {\footnotesize{}2} & {\footnotesize{}2} & {\footnotesize{}2}\tabularnewline
{\footnotesize{}$f_{\text{Sig4}}(t)$} & {\footnotesize{}3} & {\footnotesize{}3} & {\footnotesize{}3}\tabularnewline
{\footnotesize{}$f_{\text{Sig5}}(t)$} & {\footnotesize{}3} & {\footnotesize{}4} & {\footnotesize{}3}\tabularnewline
{\footnotesize{}$f_{\text{Sig6}}(t)$} & {\footnotesize{}2} & {\footnotesize{}3} & {\footnotesize{}2}\tabularnewline
\hline 
\end{tabular}{\footnotesize\par}
\end{table}

\subsection{Comparison of segmentation techniques}

\begin{figure}
\begin{centering}
\includegraphics[scale=0.6]{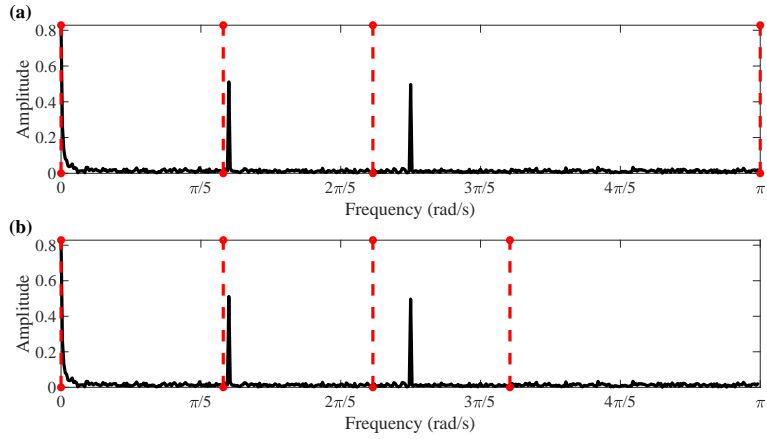}
\par\end{centering}
\caption{\label{fig:6}Segmentation results of $f_{\text{Sig1}}(t)$ by (a)
the lowest minima technique and (b) the improved segmentation technique. }

\end{figure}

\begin{figure}
\centering{}\includegraphics[scale=0.6]{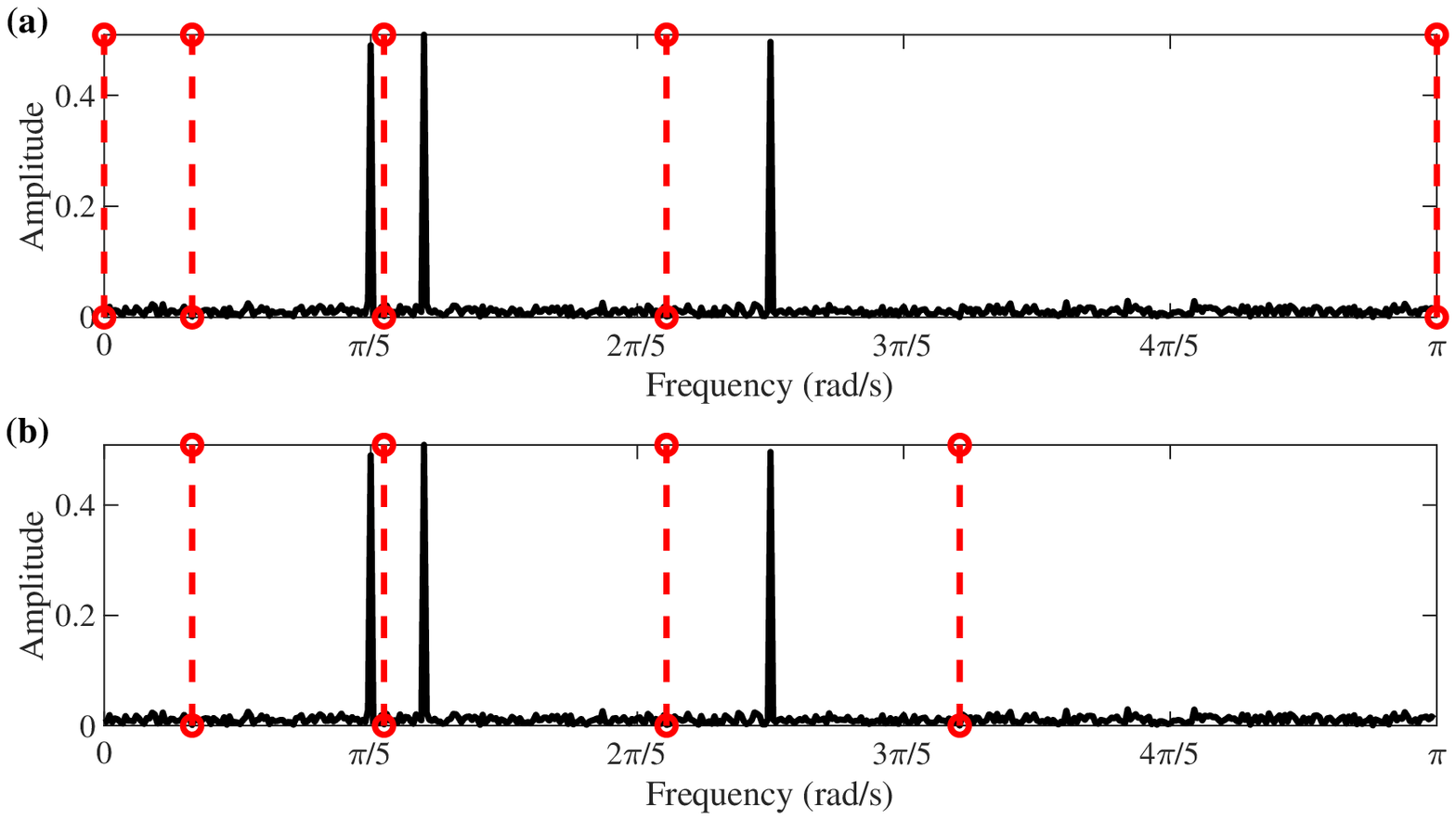}\caption{\label{fig:7}Segmentation results of $f_{\text{Sig2}}(t)$ by (a)
the lowest minima technique and (b) the improved segmentation technique.}
\end{figure}

The signals $f_{\text{Sig1}}(t)$ and $f_{\text{Sig2}}(t)$ are expressed
by

\begin{equation}
f_{\text{Sig1}}(t)=6t+\cos(24\pi t)+\cos(50\pi t)+\delta(t)
\end{equation}
and

\begin{equation}
f_{\text{Sig2}}(t)=\cos(20\pi t)+\cos(24\pi t)+\cos(50\pi t)+\delta(t)
\end{equation}
where $\delta(t)$ is a random white-noise such that $f_{\text{Sig1}}(t)$
and $f_{\text{Sig2}}(t)$ have signal-to-noise-ratios of 10 dB. Segmentation
results of $f_{\text{Sig1}}(t)$ by the lowest minima and improved
segmentation techniques are shown in Fig. \ref{fig:6}: the first
two segments by the two techniques are the same, but the last segment
in the improved segmentation technique has a narrower frequency range
than that by the lowest minima technique. Therefore, the improved
segmentation technique can alleviate the effect of noise on the decomposed
component associated with the last segment. Segmentation results of
$f_{\text{Sig2}}(t)$ by the lowest minima and improved segmentation
techniques are shown in Fig. \ref{fig:7}. The first segment by the
lowest minima technique, shown in Fig. \ref{fig:7}(a), can be considered
trivial as it does not contain a meaningful Fourier spectrum component
and its associated decomposed component consists of noise only. On
the other hand, the trivial segment is excluded in segmentation results
by the improved segmentation technique, shown in Fig. \ref{fig:7}(b),
and its first resulting segment contains a meaningful Fourier spectrum
component. Segmentation results of the last segments by the two techniques
are similar to those of $f_{\text{Sig1}}(t)$: the decomposed component
associated with the improved segmentation technique has a lower level
of noise than that by the lowest minima technique. 

\subsection{Non-stationary multimode signals}

\begin{figure}
\centering{}\includegraphics[scale=0.6]{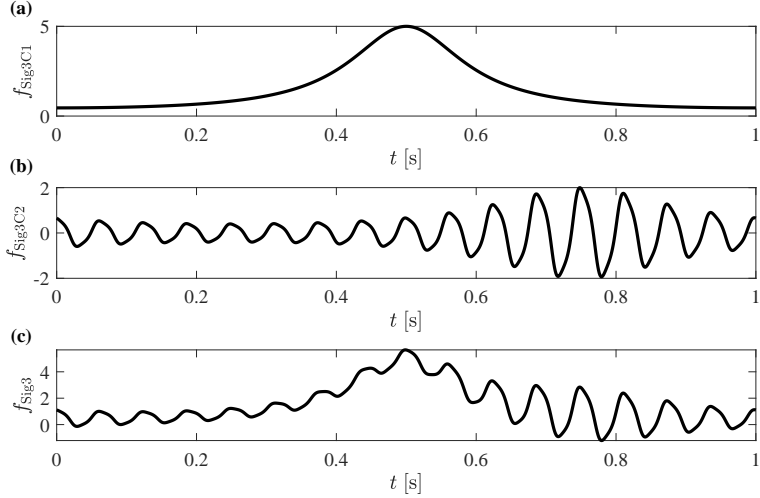}\caption{\label{fig:8}(a) Modes $f_{\text{Sig3C1}}$, (b) $f_{\text{Sig3C2}}$
and (c) the non-stationary signal $f_{\text{Sig3}}(t)$ that consists
of the two modes in (a) and (b) as expressed in Eq. (\ref{eq:34}).}
\end{figure}

The non-stationary multimode signal $f_{\text{Sig3}}(t)$ is expressed
by

\begin{equation}
\begin{cases}
f_{\text{Sig3C1}}(t) & =\frac{1}{1.2+\text{\ensuremath{\cos(2\pi t)}}}\\
f_{\text{Sig3C2}}(t) & =\frac{\cos(32\pi t+0.2\cos(64\pi t))}{1.5+\sin(2\pi t)}\\
f_{\text{Sig3}}(t) & =f_{\text{Sig3C1}}(t)+f_{\text{Sig3C2}}(t)
\end{cases}\label{eq:34}
\end{equation}
The signal $f_{\text{Sig3}}(t)$ consists of two modes $f_{\text{Sig3C1}}$and
$f_{\text{Sig3C2}}$ in Eq. (\ref{eq:34}), which are shown in Figs.
\ref{fig:8}(a) and (b) \citep{hou2011adaptive}, and $f_{\text{Sig3}}(t)$
is similar to the expression of a solution to Duffing equation \citep{huang1998empirical}.
The signal $f_{\text{Sig3}}(t)$ is sampled at a frequency of 1000
Hz for one second and shown in Fig. \ref{fig:8}(c).

\begin{figure}
\subfloat{\includegraphics[scale=0.6]{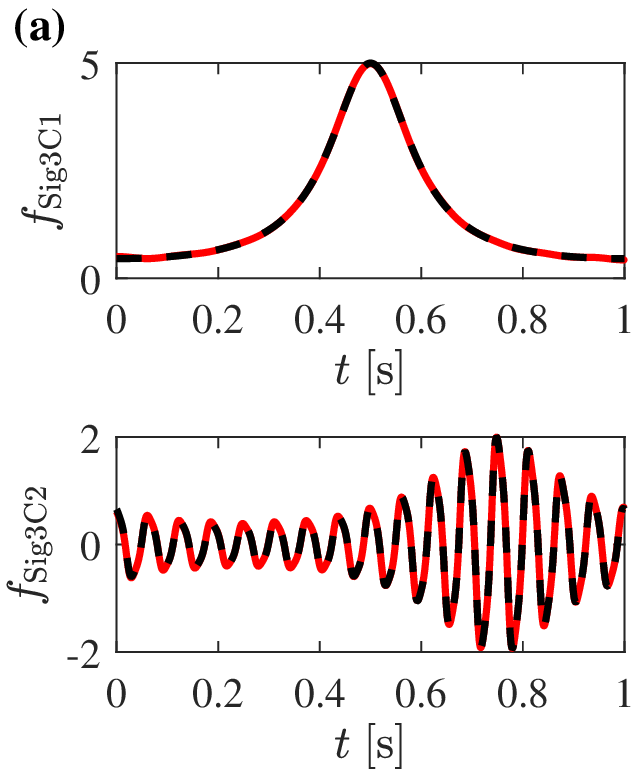}}\hfill{}\subfloat{\includegraphics[scale=0.6]{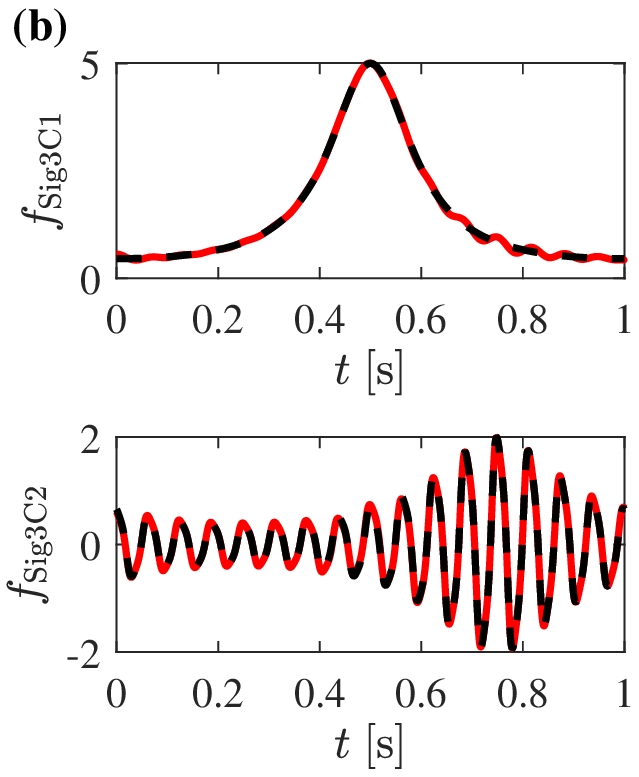}}\hfill{}\subfloat{\includegraphics[scale=0.6]{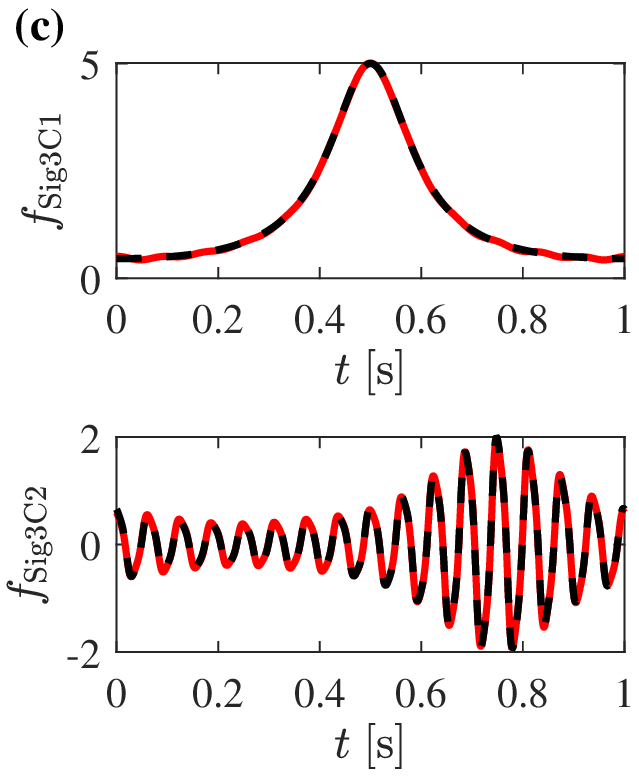}}

\subfloat{\includegraphics[scale=0.6]{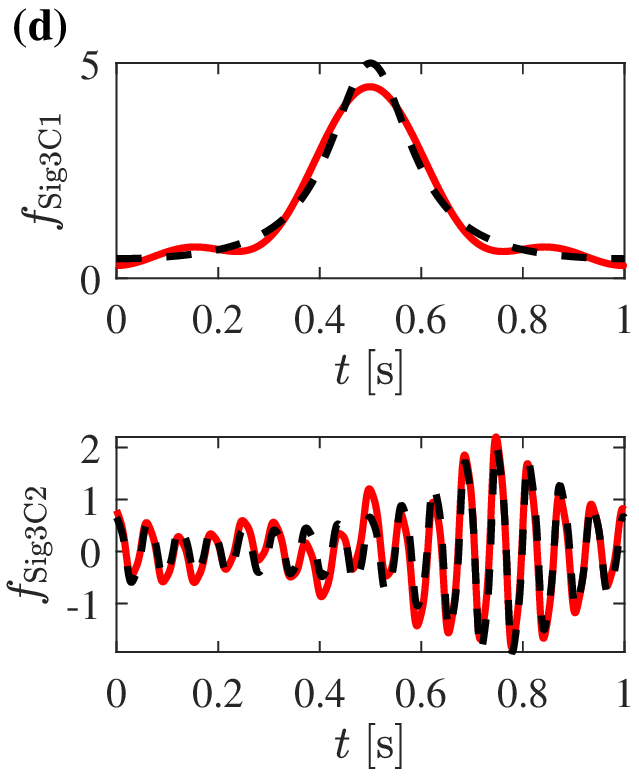}}\hfill{}\subfloat{\includegraphics[scale=0.6]{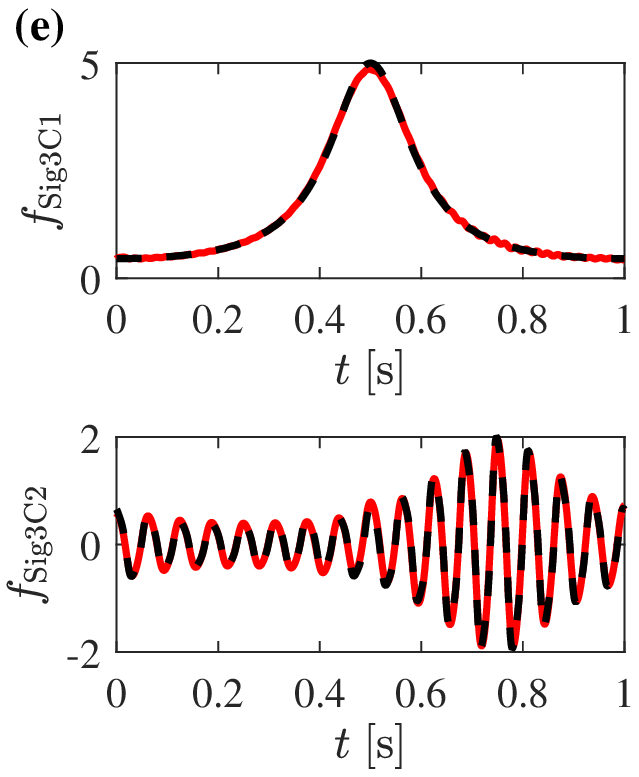}}\hfill{}\subfloat{\includegraphics[scale=0.6]{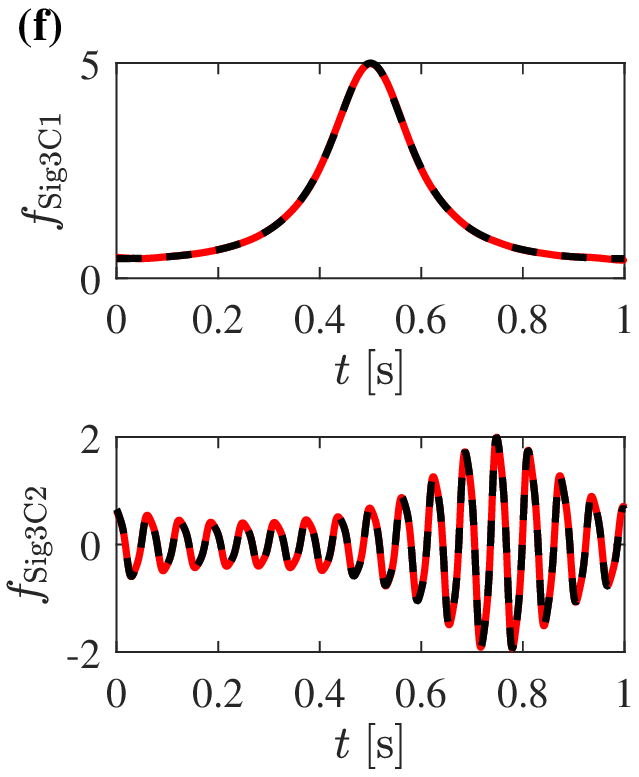}}

\caption{\label{fig:9}Comparisons between the analytical components of $f_{\text{Sig3}}(t)$
(dashed lines) in Eq. (\ref{eq:34}) and decomposition results of
$f_{\text{Sig3}}(t)$ (solid line) by the (a) EFD, (b) EWT, (c) FDM-LTH,
(d) FDM-HTL, (e) VMD, and (f) EMD.}
\end{figure}

Decomposition results of $f_{\text{Sig3}}(t)$ by the EFD, EWT, FDM-LTH,
FDM-HTL, VMD, and EMD are shown in Fig. \ref{fig:9}. Root-mean-square
errors (RMSEs) between the decomposition results and analytical ones
are calculated by

\begin{equation}
\text{RMSE=\ensuremath{\sqrt{\frac{1}{R}\sum_{r=1}^{R}\left|y_{r}-\tilde{y}_{r}\right|^{2}}}}\label{eq:35}
\end{equation}
where $y_{r}$ is the analytical component at the $r$-th discrete
instant, and $\tilde{y}_{r}$ is the corresponding component at the
$r$-th discrete instant obtained by a decomposition method. RMSEs
associated with the aforementioned decomposition methods are calculated
and listed in Table \ref{tab:2}. For $f_{\text{Sig3C1}}$, it can
be seen that the RMSE associated with the EMD is the smallest and
that associated with the EFD is the second smallest. While the RMSEs
associated with the EWT, VMD and FDM-LTH are relatively small, that
associated with the FDM-HTL is large. For $f_{\text{Sig3C2}}$, results
similar to $f_{\text{Sig3C1}}$ can be observed: RMSEs associated
with the EMD and EFD are the smallest, and the RMSEs of the EWT, VMD
and FDM-LTH are relatively small. In addition, the RMSE associated
with the FDM-HTL is also large. It is indicated that the EFD can accurately
decompose the non-stationary multimode signal. The inconsistency of
decomposition results by the FDM-LTH and FDM-HTL is verified.

\begin{table}
\caption{\label{tab:2}Results of the RMSEs for $f_{\text{Sig3}}(t)$.}

\centering{}{\footnotesize{}}%
\begin{tabular}{ccccccc}
\hline 
\multirow{2}{*}{{\footnotesize{}Component}} & \multicolumn{6}{c}{{\footnotesize{}Decomposition method}}\tabularnewline
\cline{2-7} \cline{3-7} \cline{4-7} \cline{5-7} \cline{6-7} \cline{7-7} 
 & {\footnotesize{}EFD} & {\footnotesize{}EWT} & {\footnotesize{}FDM-LTH} & {\footnotesize{}FDM-HTL} & {\footnotesize{}VMD} & {\footnotesize{}EMD}\tabularnewline
\hline 
{\footnotesize{}$f_{\text{Sig3C1}}$} & {\footnotesize{}$1.12\times10^{-2}$} & {\footnotesize{}$4.19\times10^{-2}$} & {\footnotesize{}$2.43\times10^{-2}$} & {\footnotesize{}$2.11\times10^{-1}$} & {\footnotesize{}$3.74\times10^{-2}$} & {\footnotesize{}$9.54\times10^{-3}$}\tabularnewline
{\footnotesize{}$f_{\text{Sig3C2}}$} & {\footnotesize{}$9.85\times10^{-3}$} & {\footnotesize{}$2.12\times10^{-2}$} & {\footnotesize{}$7.89\times10^{-2}$} & {\footnotesize{}$2.24\times10^{-1}$} & {\footnotesize{}$5.41\times10^{-2}$} & {\footnotesize{}$9.54\times10^{-3}$}\tabularnewline
\hline 
\end{tabular}{\footnotesize\par}
\end{table}

\begin{figure}
\centering{}\includegraphics[scale=0.6]{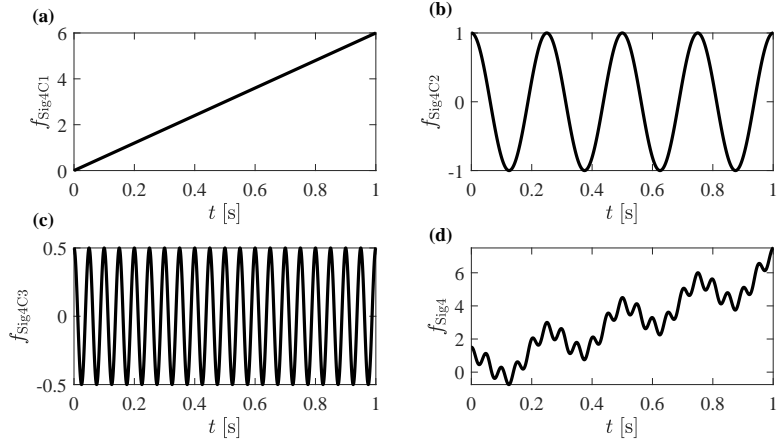}\caption{\label{fig:10}(a) Modes $f_{\text{Sig4C1}}$, (b) $f_{\text{Sig4C2}}$,
(c) $f_{\text{Sig4C3}}$ and (d) the non-stationary signal $f_{\text{Sig4}}(t)$
that consists of the three modes in (a), (b) and (c) as expressed
in Eq. (\ref{eq:36}).}
\end{figure}

Another non-stationary multimode signal $f_{\text{Sig4}}(t)$ is expressed
by \citep{hou2011adaptive}

\begin{equation}
\begin{cases}
f_{\text{Sig4C1}}(t) & =6t\\
f_{\text{Sig4C2}}(t) & =\cos(8\pi t)\\
f_{\text{Sig4C3}}(t) & =0.5\cos(40\pi t)\\
f_{\text{Sig4}}(t) & =f_{\text{Sig4C1}}(t)+f_{\text{Sig4C2}}(t)+f_{\text{Sig4C3}}(t)
\end{cases}\label{eq:36}
\end{equation}
The signal $f_{\text{Sig4}}(t)$ consists of three modes: one mode
$f_{\text{Sig4C1}}$ with a monotonically increasing amplitude as
shown in Fig. \ref{fig:10}(a) and two modes $f_{\text{Sig4C2}}$
and $f_{\text{Sig4C3}}$ with sinusoidal amplitudes as shown in Figs.
\ref{fig:10}(b) and (c), respectively. The signal $f_{\text{Sig4}}(t)$
is sampled at a frequency of 1000 Hz for one second and shown in Fig.
\ref{fig:10}(d).

\begin{figure}
\subfloat{\includegraphics[scale=0.6]{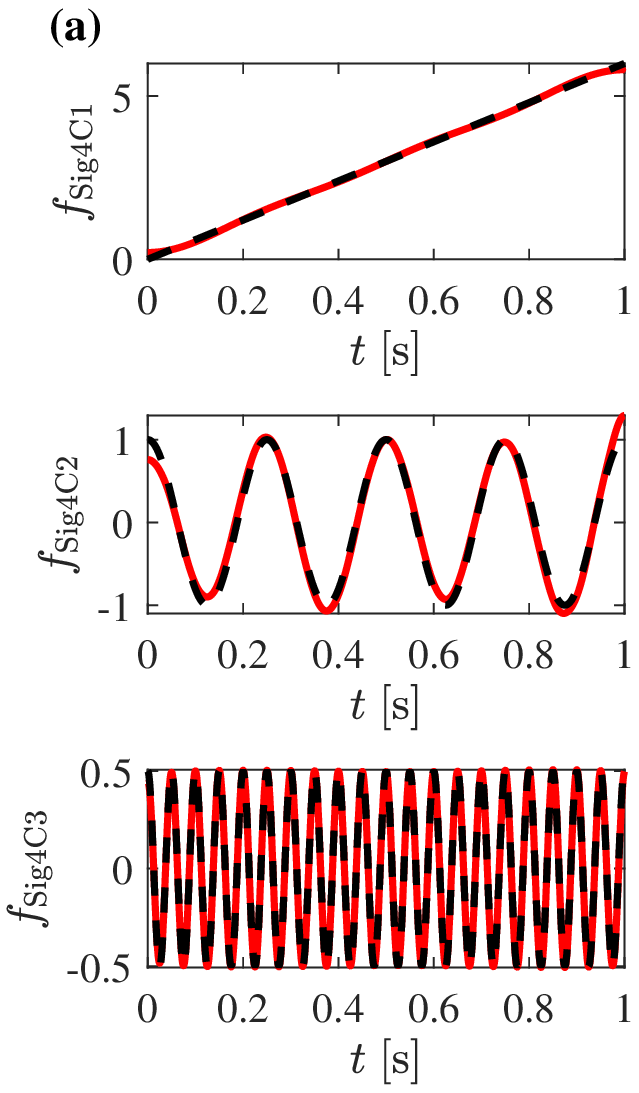}}\hfill{}\subfloat{\includegraphics[scale=0.6]{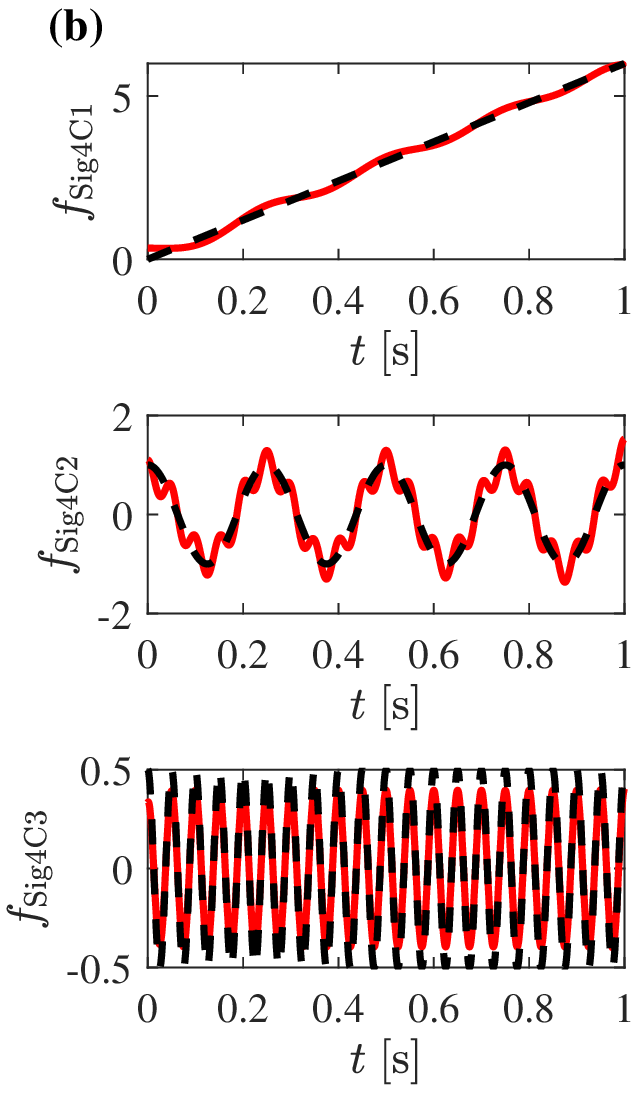}}\hfill{}\subfloat{\includegraphics[scale=0.6]{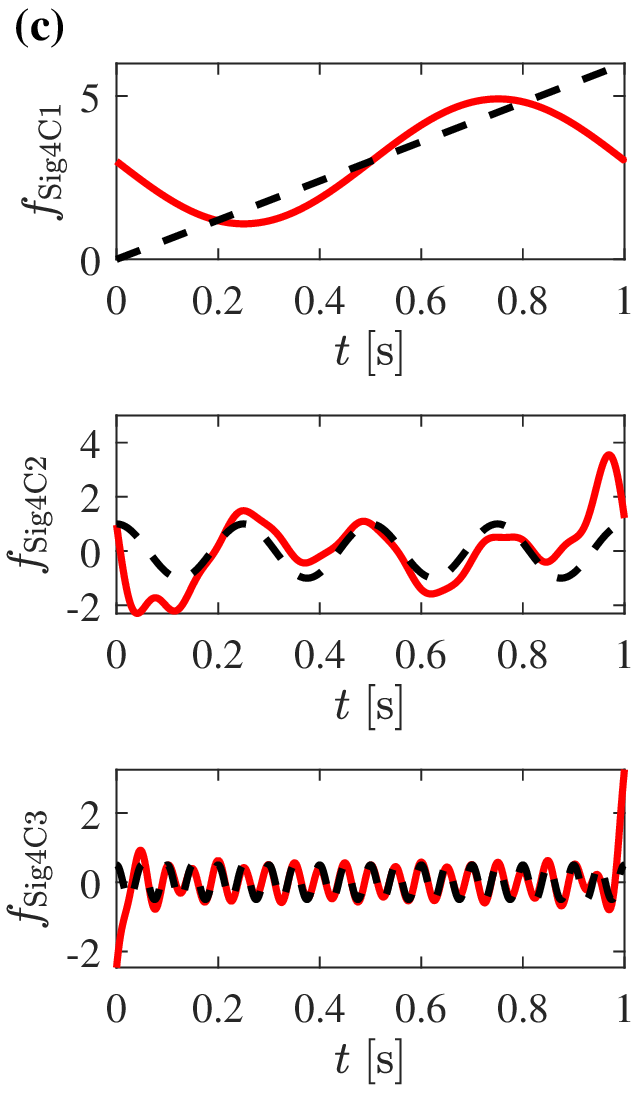}}

\subfloat{\includegraphics[scale=0.6]{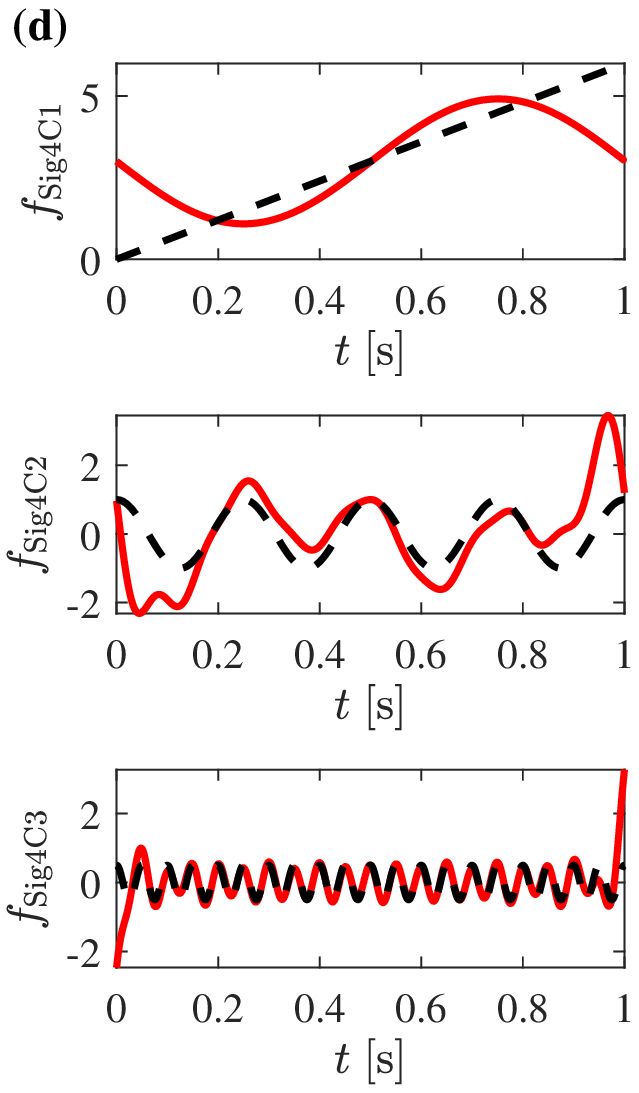}}\hfill{}\subfloat{\includegraphics[scale=0.6]{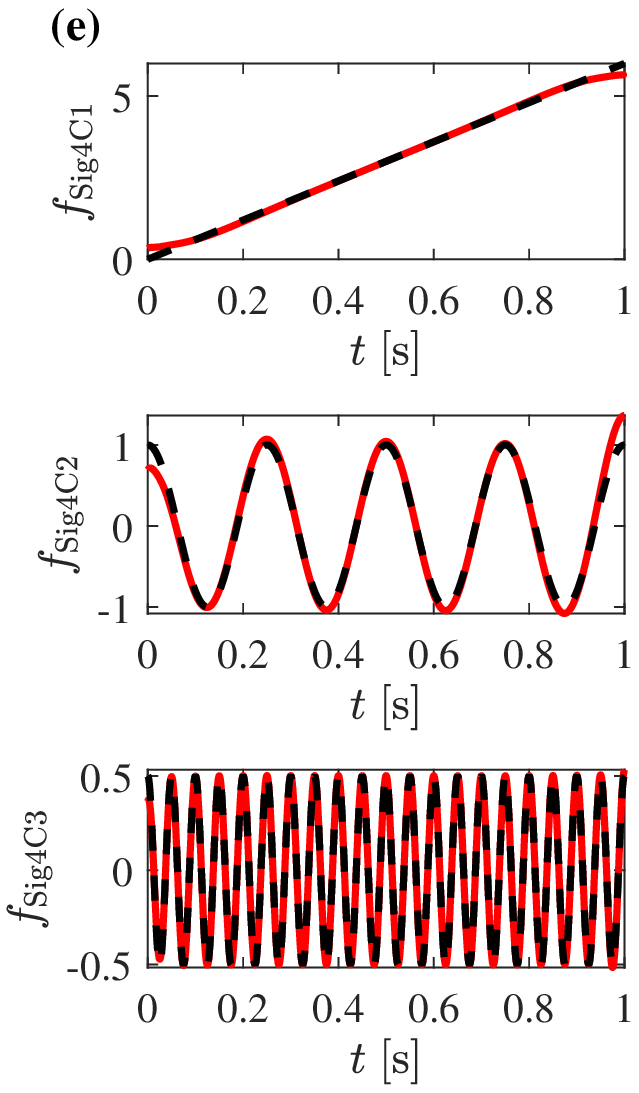}}\hfill{}\subfloat{\includegraphics[scale=0.6]{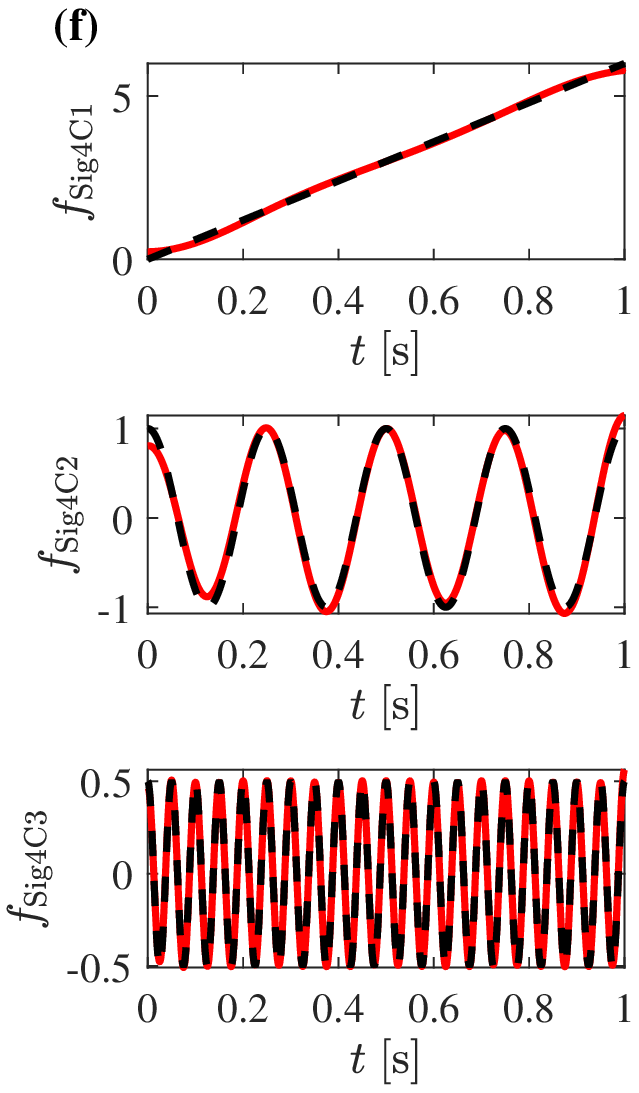}}\caption{\label{fig:11}Comparisons between the analytical components of $f_{\text{Sig4}}(t)$
(dashed lines) in Eq. (\ref{eq:36}) and decomposition results of
$f_{\text{Sig4}}(t)$ (solid line) by the (a) EFD, (b) EWT, (c) FDM-LTH,
(d) FDM-HTL, (e) VMD, and (f) EMD.}
\end{figure}

The EFD, EWT, FDM-LTH, FDM-HTL, VMD and EMD are employed to decompose
the sampled $f_{\text{Sig4}}(t)$. Their results are compared with
the analytical ones as shown in Fig. \ref{fig:11} and corresponding
RMSEs in Eq. (\ref{eq:35}) are calculated and listed in Table \ref{tab:3}.
For $f_{\text{Sig4C1}}$, the RMSE associated with the EFD is the
smallest, and those associated with the EWT, VMD and EMD are relatively
small. RMSEs associated with the FDM-LTH and FDM-HTL are large. For
$f_{\text{Sig4C2}}$, RMSE associated the EMD is the smallest and
those associated with the EFD and VMD are slightly larger than that
associated with the EMD. The RMSEs associated with the EWT, FDM-LTH
and FDM-HTL are large and that associated with the FDM is the largest.
For $f_{\text{Sig4C3}}$, the RMSE associated with the EFD is the
smallest while those associated with the VMD and EMD are slightly
larger than that associated with the EFD. Similar to the observations
for $f_{\text{Sig4C2}}$, the RMSEs associated with the EWT, FDM-TLH
and FDM-HTL are larger than others. It is indicated again that the
EFD can accurately decompose a nonstationary multimode signal. In
addition, it is shown that both the FDM-LTH and FDM-HTL can yield
inaccurate decomposition results.

\begin{table}
\caption{\label{tab:3}Results of the RMSEs for $f_{\text{Sig4}}(t)$.}

\centering{}{\footnotesize{}}%
\begin{tabular}{ccccccc}
\hline 
\multirow{2}{*}{{\footnotesize{}Component}} & \multicolumn{6}{c}{{\footnotesize{}Decomposition method}}\tabularnewline
\cline{2-7} \cline{3-7} \cline{4-7} \cline{5-7} \cline{6-7} \cline{7-7} 
 & {\footnotesize{}EFD} & {\footnotesize{}EWT} & {\footnotesize{}FDM-LTH} & {\footnotesize{}FDM-HTL} & {\footnotesize{}VMD} & {\footnotesize{}EMD}\tabularnewline
\hline 
{\footnotesize{}$f_{\text{Sig4C1}}$} & {\footnotesize{}$4.67\times10^{-2}$} & {\footnotesize{}$1.07\times10^{-1}$} & {\footnotesize{}$1.08\times10^{0}$ } & {\footnotesize{}$1.08\times10^{0}$ } & {\footnotesize{}$8.40\times10^{-2}$} & {\footnotesize{}$6.46\times10^{-2}$ }\tabularnewline
{\footnotesize{}$f_{\text{Sig4C2}}$} & {\footnotesize{}$8.33\times10^{-2}$} & {\footnotesize{}$2.23\times10^{-1}$} & {\footnotesize{}$1.02\times10^{0}$ } & {\footnotesize{}$1.02\times10^{0}$ } & {\footnotesize{}$9.14\times10^{-2}$} & {\footnotesize{}$6.38\times10^{-2}$ }\tabularnewline
{\footnotesize{}$f_{\text{Sig4C3}}$} & {\footnotesize{}$7.01\times10^{-3}$ } & {\footnotesize{}$7.43\times10^{-2}$ } & {\footnotesize{}$3.81\times10^{-1}$} & {\footnotesize{}$3.98\times10^{-1}$} & {\footnotesize{}$9.93\times10^{-3}$} & {\footnotesize{}$8.10\times10^{-3}$ }\tabularnewline
\hline 
\end{tabular}{\footnotesize\par}
\end{table}

\subsection{Closely-spaced modes}

\begin{figure}
\centering{}\includegraphics[scale=0.6]{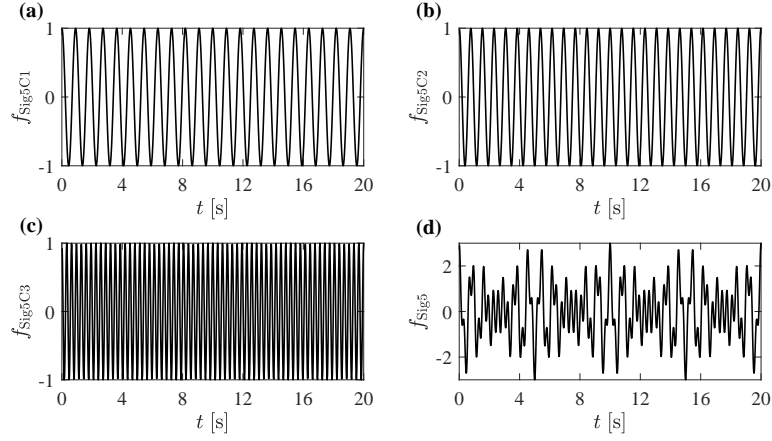}\caption{\label{fig:12}(a) Modes $f_{\text{Sig5C1}}$, (b) $f_{\text{Sig5C2}}$,
(c) $f_{\text{Sig5C3}}$ and (d) $f_{\text{Sig5}}(t)$ that consists
of the three modes in (a), (b) and (c) which are expressed in Eq.
(\ref{eq:37}).}
\end{figure}

A stationary signal $f_{\text{Sig5}}(t)$ with two closely-spaced
modes is expressed by

\begin{equation}
\begin{cases}
f_{\text{Sig5C1}}(t) & =\cos(2\text{\ensuremath{\pi\lambda_{a}t}})\\
f_{\text{Sig5C2}}(t) & =\cos(2\text{\ensuremath{\pi\lambda_{b}t}})\\
f_{\text{Sig5C3}}(t) & =\cos(2\text{\ensuremath{\pi\lambda_{c}t}})\\
f_{\text{Sig5}}(t) & =f_{\text{Sig5C1}}(t)+f_{\text{Sig5C2}}(t)+f_{\text{Sig5C3}}(t)
\end{cases}\label{eq:37}
\end{equation}
where $\lambda_{a}$=1.1 Hz, $\lambda_{b}$=1.3 Hz and $\lambda_{c}$=3.1
Hz, and it consists of a pair of closely-spaced modes shown in Figs.
\ref{fig:12}(a) and (b) and, a mode with a frequency greatly larger
than those of the closely-spaced modes is shown in Fig. \ref{fig:12}(c).
The signal $f_{\text{Sig5}}(t)$ is sampled at a frequency of 50 Hz
for 20 seconds and shown in Fig. \ref{fig:12}(d).

\begin{figure}
\subfloat{\includegraphics[scale=0.6]{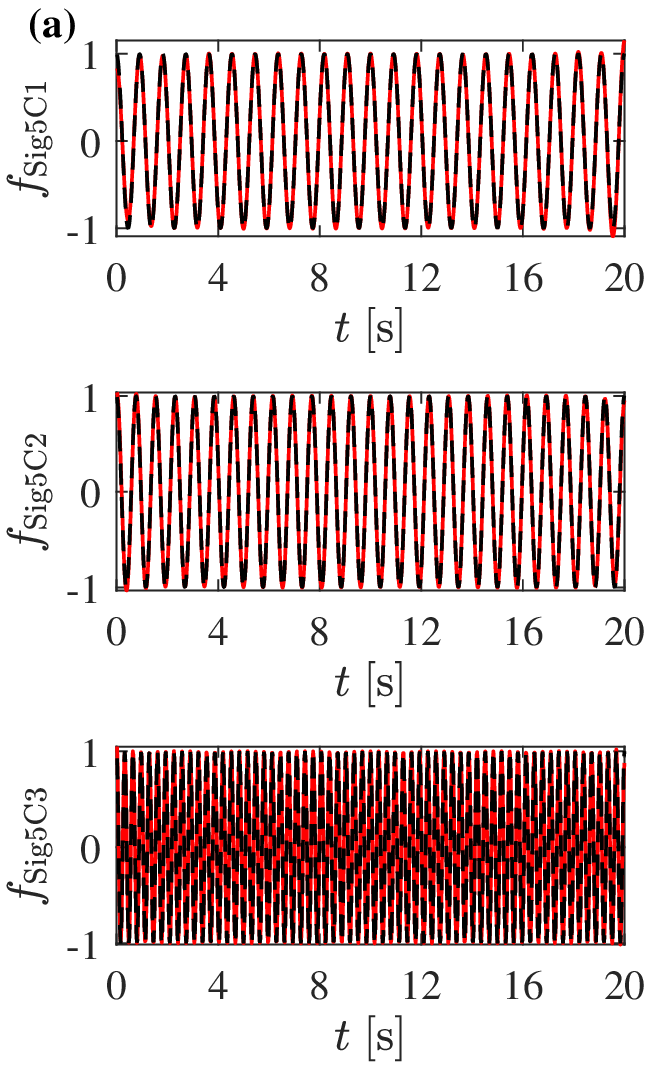}}\hfill{}\subfloat{\includegraphics[scale=0.6]{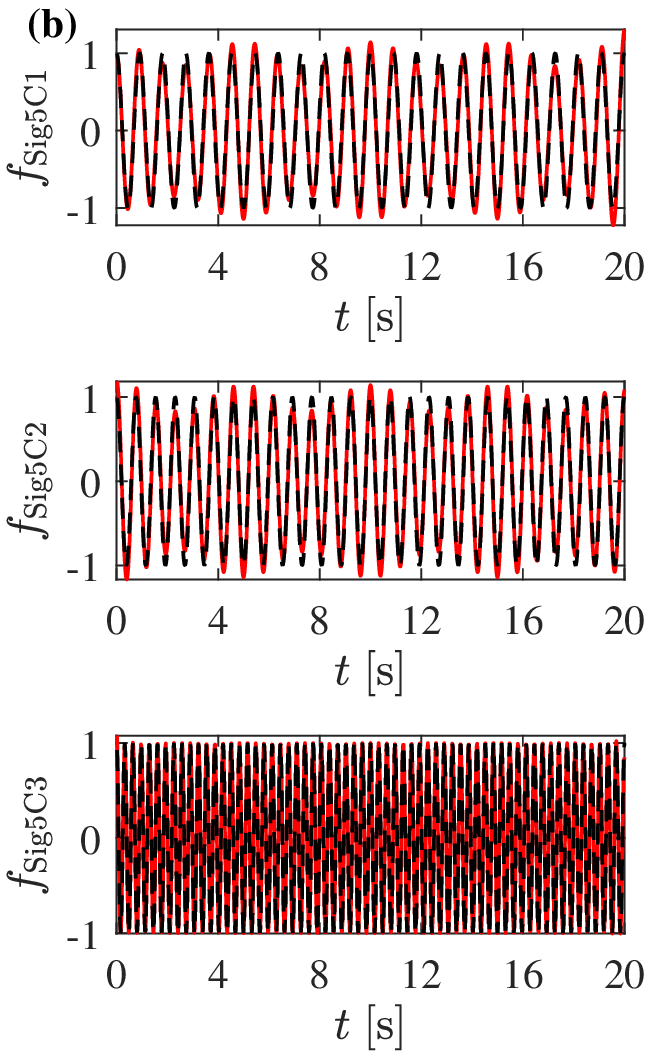}}\hfill{}\subfloat{\includegraphics[scale=0.6]{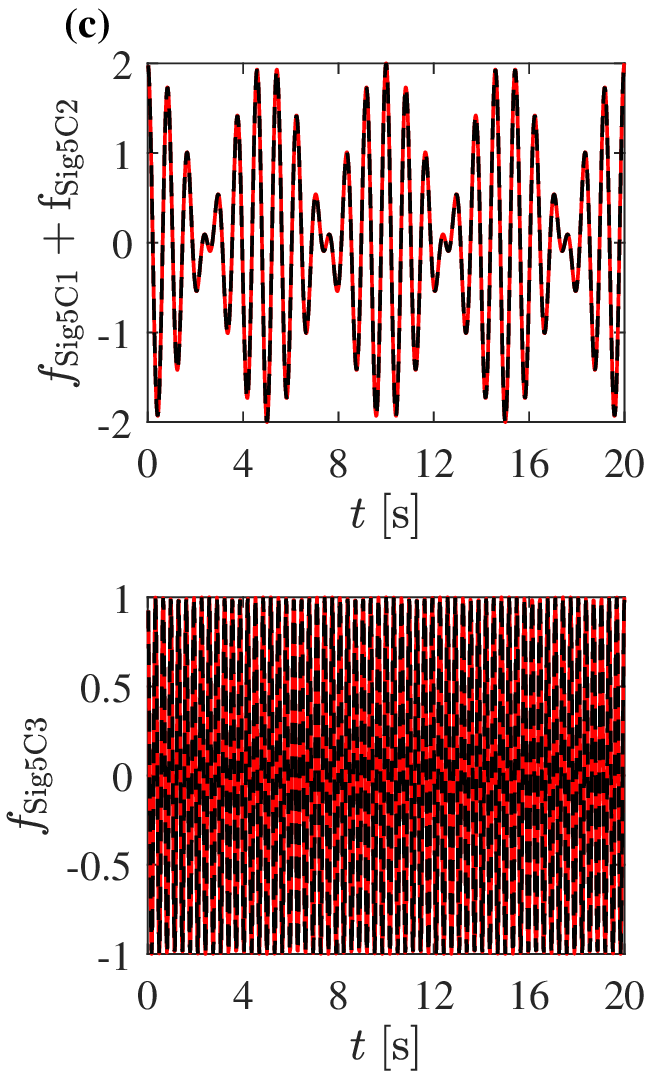}}

\subfloat{\includegraphics[scale=0.6]{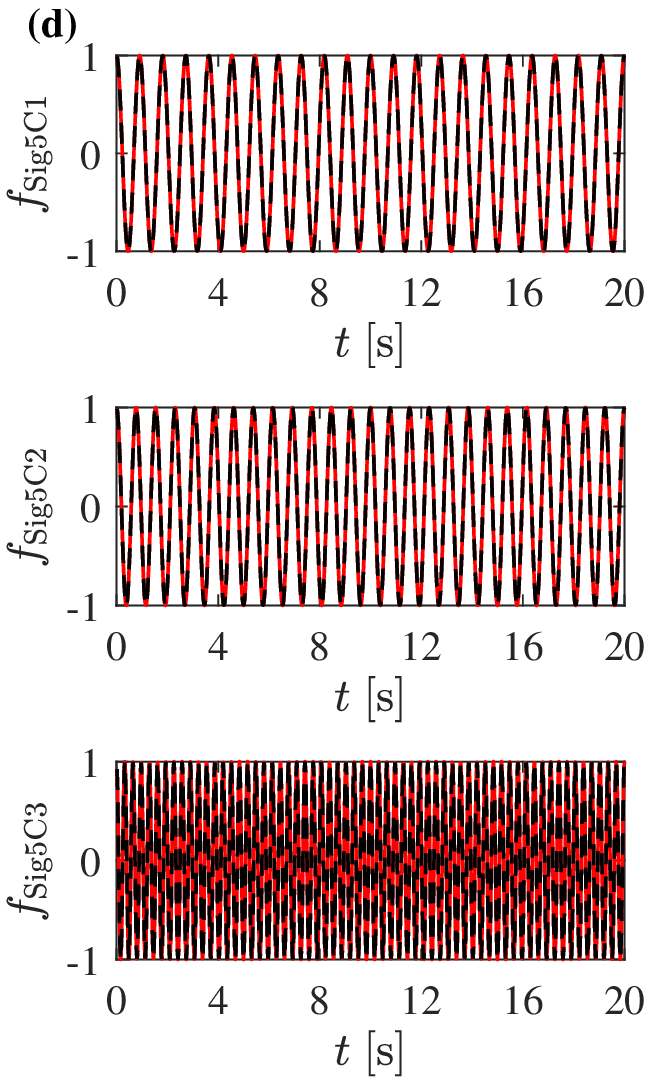}}\hfill{}\subfloat{\includegraphics[scale=0.6]{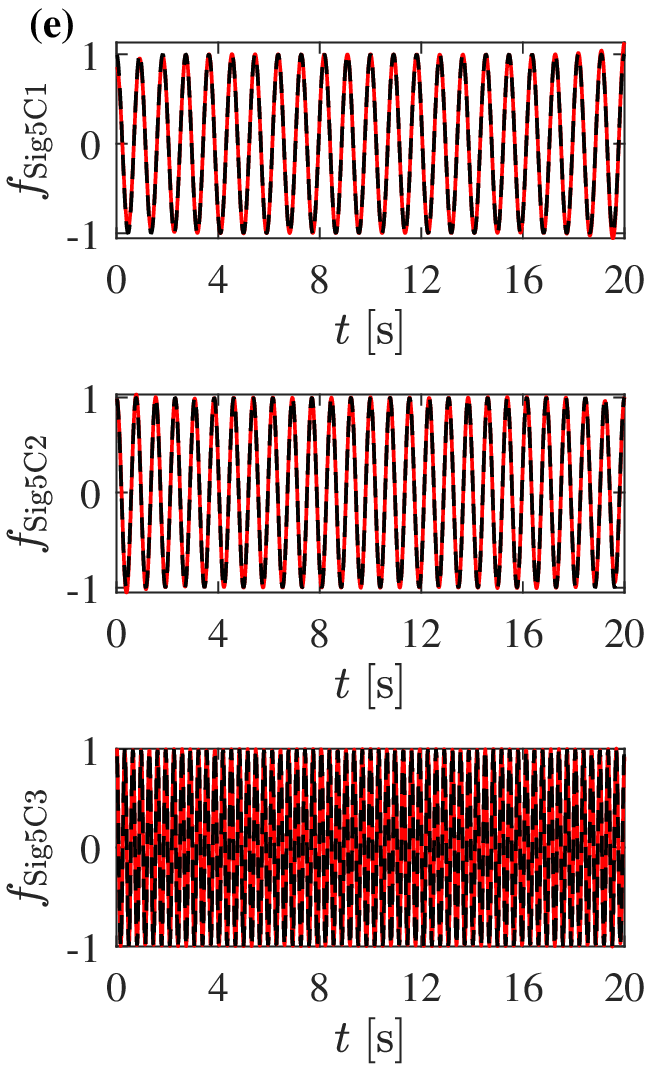}}\hfill{}\subfloat{\includegraphics[scale=0.6]{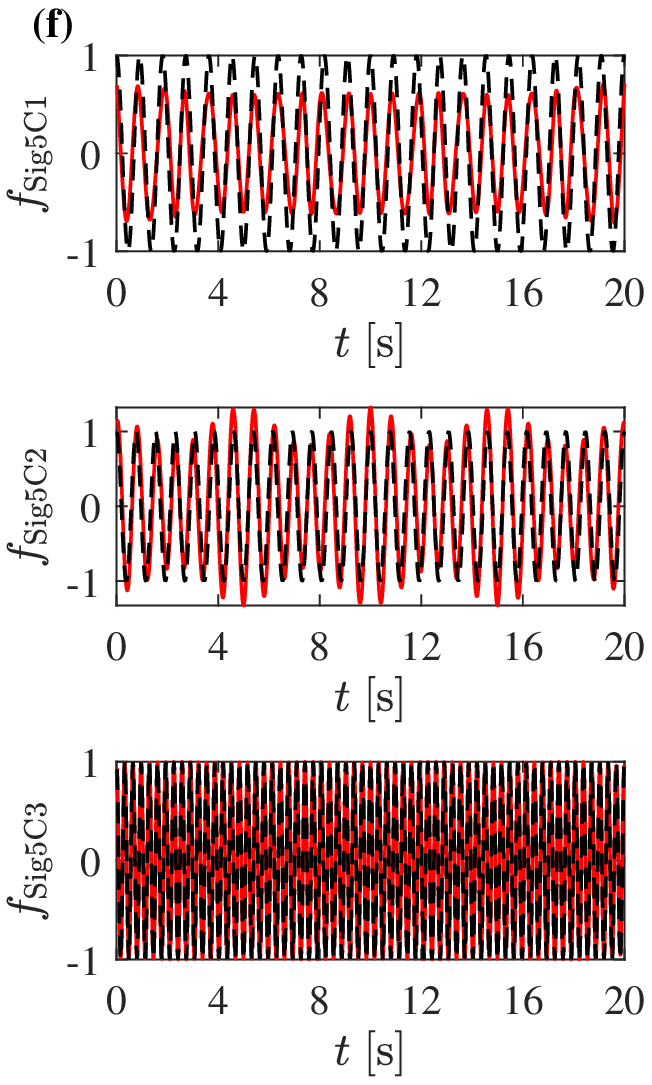}}\caption{\label{fig:13}Comparisons between the analytical components of $f_{\text{Sig5}}(t)$
(dashed lines) in Eq. (\ref{eq:37}) and decomposition results of
$f_{\text{Sig5}}(t)$ (solid line) by the (a) EFD, (b) EWT, (c) FDM-LTH,
(d) FDM-HTL, (e) VMD, and (f) EMD.}
\end{figure}

The EFD, EWT, FDM-LTH, FDM-HTL, VMD, and EMD are used to decompose
the sampled $f_{\text{Sig5}}(t)$. Their results are shown in Fig.
\ref{fig:13} and corresponding RMSEs in Eq. (\ref{eq:36}) are calculated
and listed in Table \ref{tab:4}. Note that in the decomposition result
by the FDM-LTH, only two components are obtained as $f_{\text{Sig5C1}}$
and $f_{\text{Sig5C2}}$ exist in the first component and RMSEs corresponding
to $f_{\text{Sig5C1}}$ and $f_{\text{Sig5C2}}$ are mixed as one
mode. For $f_{\text{Sig5C1}}$, the RMSE associated with the FDM-HTL
is the smallest and those associated with the VMD and EFD are relatively
small, and those associated with the EWT, FDM-LTH and EMD are large.
For $f_{\text{Sig5C2}}$, similar observations can be obtained: the
RMSE associated with the FDM-HTL is the smallest, those associated
with the VMD and EFD are relatively small, and those associated with
EWT, FDM-LTH and EMD are large. For $f_{\text{Sig5C3}}$, the RMSEs
associated with the FDM-LTH and FDM-HTL are the smallest and those
associated with other methods are relatively small. It is indicated
that the EFD can yield decomposition results for signals with closely-spaced
modes with higher accuracy than the EWT and EMD. In addition, the
inconsistency between decomposition results by the FDM-LTH and FDM-HTL
is verified again.

\begin{table}
\caption{\label{tab:4}Results of the RMSEs for $f_{\text{Sig5}}(t)$.}

\centering{}{\footnotesize{}}%
\begin{tabular}{ccccccc}
\hline 
\multirow{2}{*}{{\footnotesize{}Component}} & \multicolumn{6}{c}{{\footnotesize{}Decomposition method}}\tabularnewline
\cline{2-7} \cline{3-7} \cline{4-7} \cline{5-7} \cline{6-7} \cline{7-7} 
 & {\footnotesize{}EFD} & {\footnotesize{}EWT} & {\footnotesize{}FDM-LTH} & {\footnotesize{}FDM-HTL} & {\footnotesize{}VMD} & {\footnotesize{}EMD}\tabularnewline
\hline 
{\footnotesize{}$f_{\text{Sig4C1}}$} & {\footnotesize{}$2.20\times10^{-2}$} & {\footnotesize{}$1.11\times10^{-1}$} & {\footnotesize{}$7.07\times10^{-1}$ } & {\footnotesize{}$0.00$ } & {\footnotesize{}$1.90\times10^{-2}$} & {\footnotesize{}$7.02\times10^{-1}$ }\tabularnewline
{\footnotesize{}$f_{\text{Sig4C2}}$} & {\footnotesize{}$1.60\times10^{-2}$} & {\footnotesize{}$1.10\times10^{-1}$} & {\footnotesize{}$7.07\times10^{-1}$ } & {\footnotesize{}$0.00$ } & {\footnotesize{}$1.50\times10^{-2}$} & {\footnotesize{}$6.94\times10^{-1}$ }\tabularnewline
{\footnotesize{}$f_{\text{Sig4C3}}$} & {\footnotesize{}$1.20\times10^{-2}$ } & {\footnotesize{}$1.20\times10^{-2}$ } & {\footnotesize{}$0.00$} & {\footnotesize{}$0.00$ } & {\footnotesize{}$1.41\times10^{-2}$} & {\footnotesize{}$4.58\times10^{-2}$ }\tabularnewline
\hline 
\end{tabular}{\footnotesize\par}
\end{table}

\begin{figure}
\subfloat{\includegraphics[scale=0.6]{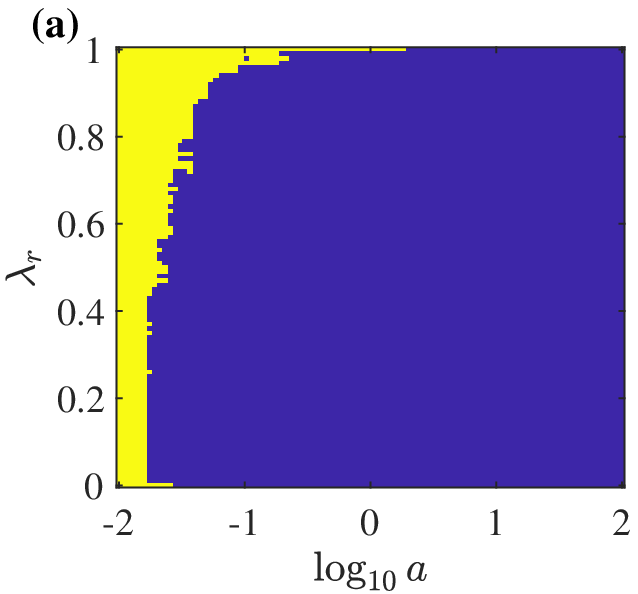}}\hfill{}\subfloat{\includegraphics[scale=0.6]{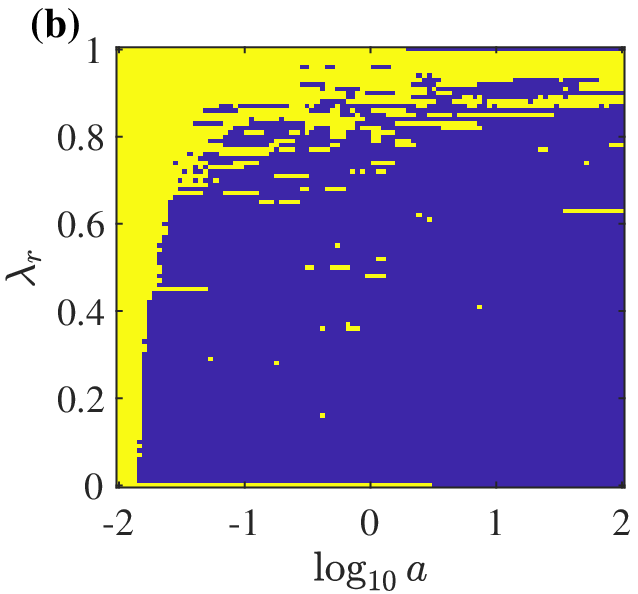}}\hfill{}\subfloat{\includegraphics[scale=0.6]{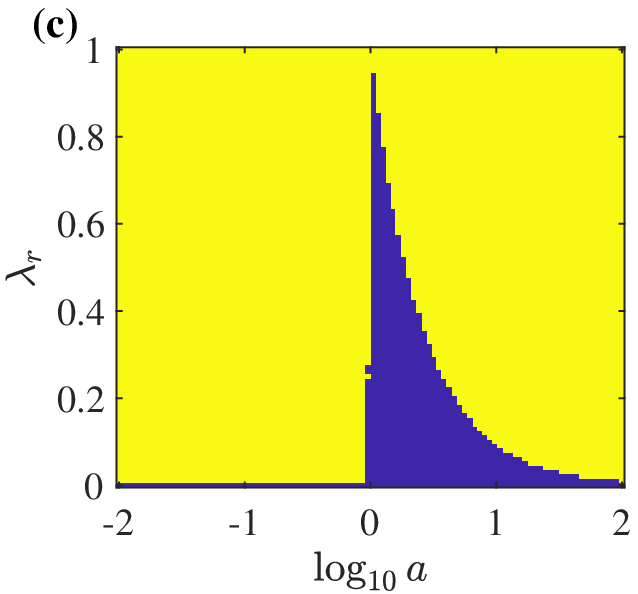}}\hfill{}

\subfloat{\includegraphics[scale=0.6]{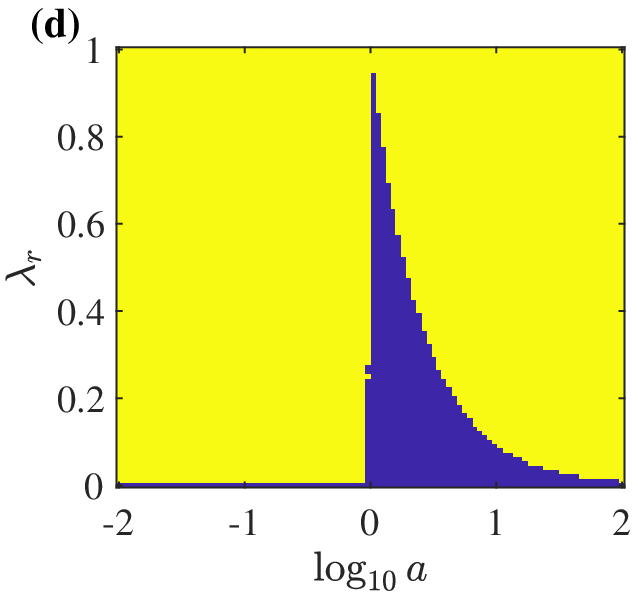}}\hfill{}\subfloat{\includegraphics[scale=0.6]{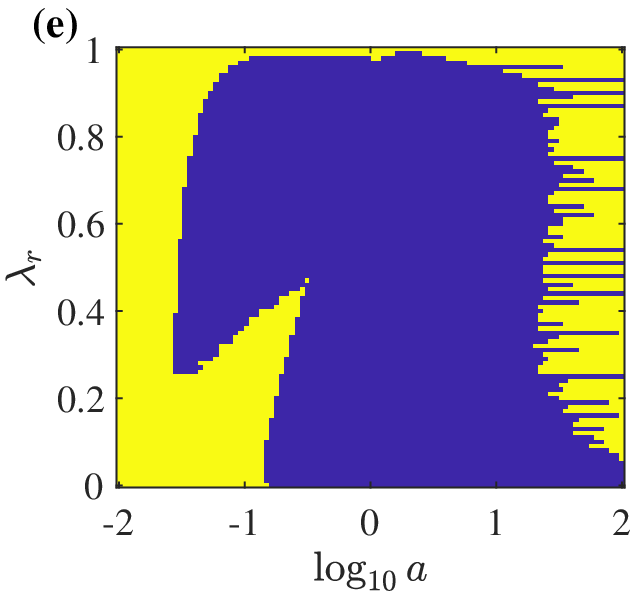}}\hfill{}\subfloat{\includegraphics[scale=0.6]{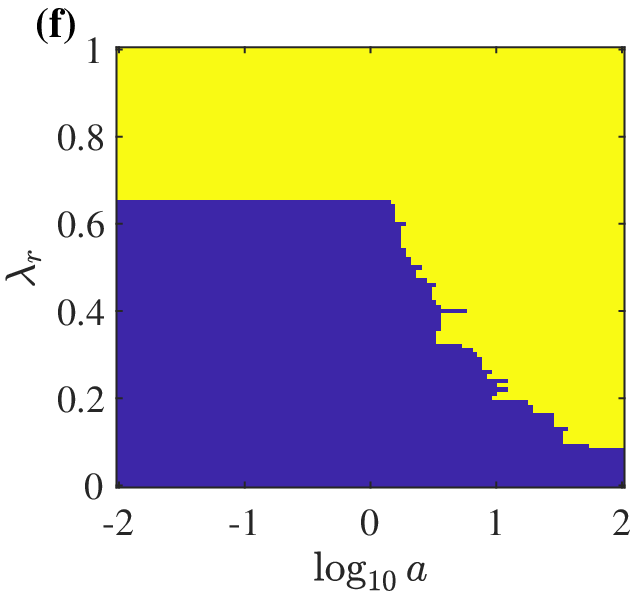}}\hfill{}\caption{\label{fig:14}Decomposition performance with respect to $(a,\lambda_{r})$
by the (a) EFD, (b) EWT, (c) FDM-LTH, (d) FDM-HTL, (e) VMD and (f)
EMD.}
\end{figure}

Another stationary signal with two modes \citep{rilling2007one} denoted
by $f_{\text{Sig6}}(t)$ is constructed to further compare performances
of the different decomposition methods for signals with closely-spaced
modes, which is expressed by

\begin{equation}
\begin{cases}
f_{\text{Sig6C1}}(t) & =\cos(2\text{\ensuremath{\pi t}})\\
f_{\text{Sig6C2}}(t) & =a\cos(2\text{\ensuremath{\pi\lambda_{r}t}})\\
f_{\text{Sig6}}(t) & =f_{\text{Sig6C1}}(t)+f_{\text{Sig6C2}}(t)
\end{cases}\label{eq:38}
\end{equation}
where $a$ and $\lambda_{r}$ denote a ratio between the amplitudes
of $f_{\text{Sig6C2}}(t)$ and $f_{\text{Sig6C1}}(t)$ and that between
the frequencies of $f_{\text{Sig6C2}}(t)$ and $f_{\text{Sig6C1}}(t)$,
respectively; 0.01 \ensuremath{\le} $a$ \ensuremath{\le} 100 and
0.01 \ensuremath{\le} $\lambda_{r}$ \ensuremath{\le} 1. When $\lambda_{r}$
approaches to 1, $f_{\text{Sig6C2}}(t)$ and $f_{\text{Sig6C1}}(t)$
become closely-spaced modes. The signal $f_{\text{Sig6}}(t)$ is sampled
at a frequency of 10 Hz for 300 seconds. The EFD, EWT, FDM-LTH, FDM-HLT,
VMD and EMD are deployed to decompose $f_{\text{Sig6}}(t)$. A two-dimensional
binary quantity $Q(a,\lambda_{r})$ is used to measure the decomposition
performance \citep{rilling2007one} of the different methods for $f_{\text{Sig6}}(t)$
with different values of $a$ and $\lambda_{r}$, and it is expressed
by 

\begin{equation}
Q(a,\lambda_{r})=\begin{cases}
0 & \textrm{if}\frac{\left\Vert \text{C}1-f_{\text{Sig6C1}}\right\Vert _{2}}{\left\Vert f_{\text{Sig6C2}}\right\Vert _{2}}\leq\varepsilon\\
1 & \textrm{if}\frac{\left\Vert \text{C}1-f_{\text{Sig6C1}}\right\Vert _{2}}{\left\Vert f_{\text{Sig6C2}}\right\Vert _{2}}>\varepsilon
\end{cases}\label{eq:39}
\end{equation}
where C1 is the decomposed component by a decomposition method corresponding
to $f_{\text{Sig6C1}}(t)$ and $\varepsilon$ is the threshold of
$Q$. A zero value and a unit value of $Q$ indicate an acceptable
decomposition result and an unacceptable decomposition one, respectively,
and the value of $\varepsilon$ is chosen to be 0.5 in this study,
which was also the cases in Refs. \citep{chen2019adaptive,li2017time,rilling2007one}.
Resulting $Q(a,\lambda_{r})$ corresponding to the six methods are
shown in Fig. \ref{fig:14}, where the colors of blue and yellow correspond
to $Q$ values of 0 and 1, respectively. It can be seen that the yellow
area corresponding to the EFD is the smallest among the six $Q$ results.
Even as $\lambda_{r}$ approaches to 1, $f_{\text{Sig6}}(t)$ can
still be well decomposed. However, the decomposition by the EFD is
affected when $a$ approaches to 0.01. Further, the yellow area corresponding
to the EWT is the second smallest but its decomposition performance
is affected when $a$ approaches to 0.01 and $\lambda_{r}$ is larger
than 0.8. The yellow area corresponding to the VMD is the third smallest.
Similar to Q corresponding to the EFD, as $\lambda_{r}$ approaches
to 1, $f_{\text{Sig6}}(t)$ can still be well decomposed. However,
the decomposition performance associated with the VMD is affected
when $a$ is close to 0.01 and 100. The yellow area corresponding
to the EMD is the fourth smallest. The EMD cannot decompose $f_{\text{Sig6}}(t)$,
when $\lambda_{r}$ is larger than 0.65 for all $a$. In addition,
worse decomposition results are obtained when $a$ approaches to 100.
For the FDM-LTH and FDM-HTL, decomposition performances are almost
the same but the worst among the six methods. Their decomposition
results are greatly affected by the value of $a$. They can hardly
decompose $f_{\text{Sig6}}(t)$, when $a$ is smaller than 1 for $\lambda_{r}\geq0.01$.
Based on the observations, it is indicated that the EFD can rrobustly
and accurately decompose $f_{\text{Sig6}}(t)$ as its decomposition
results are the most accurate even when $f_{\text{Sig6}}(t)$ becomes
a signal with closely-spaced modes, and both the FDM-LTH and FDM-HTL
can yield inaccurate decomposition results for $f_{\text{Sig6}}(t)$. 

\subsection{TFR}

\begin{figure}
\subfloat{\centering{}\includegraphics[scale=0.6]{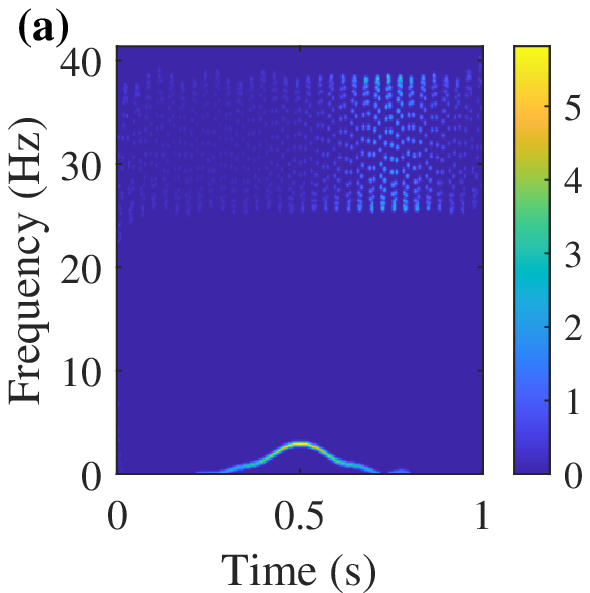}}\hfill{}\subfloat{\centering{}\includegraphics[scale=0.6]{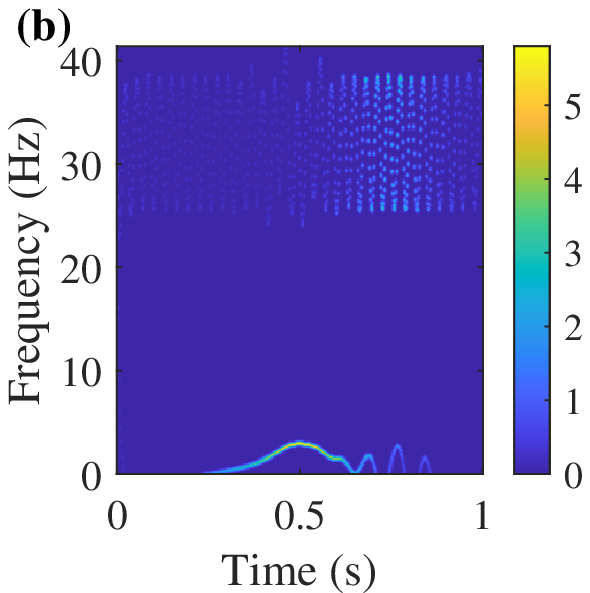}}\hfill{}\subfloat{\centering{}\includegraphics[scale=0.6]{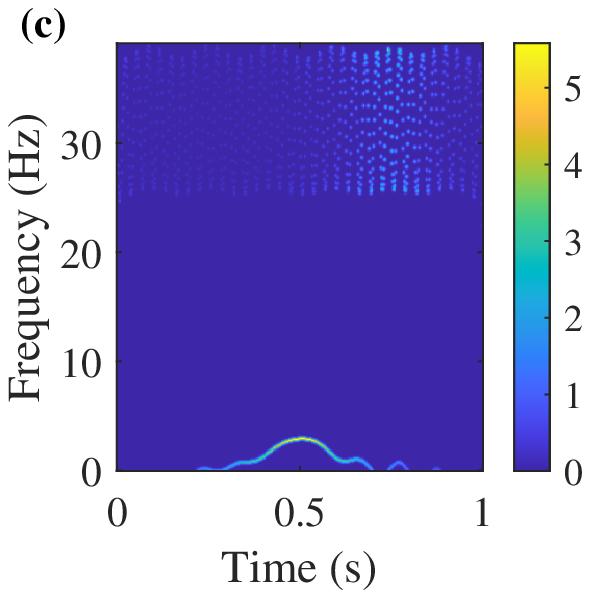}}\hfill{}
\begin{centering}
\subfloat{\centering{}\includegraphics[scale=0.6]{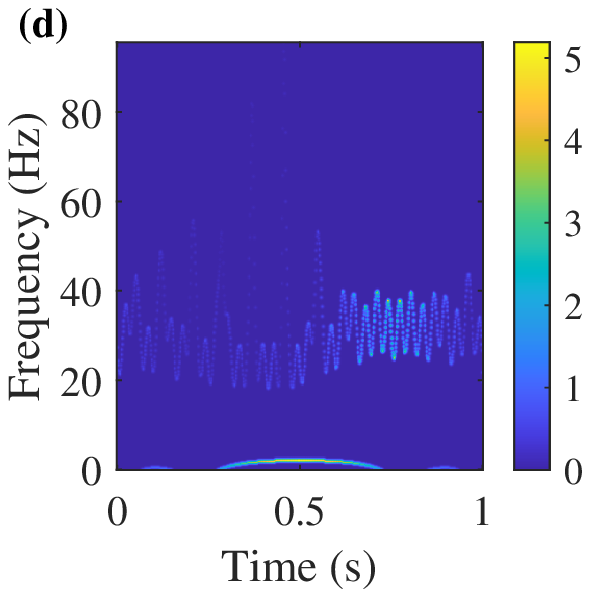}}\hfill{}\subfloat{\centering{}\includegraphics[scale=0.6]{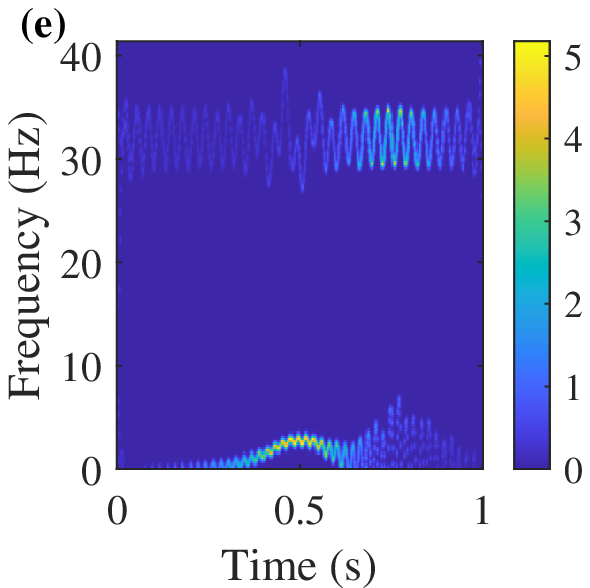}}\hfill{}\includegraphics[scale=0.6]{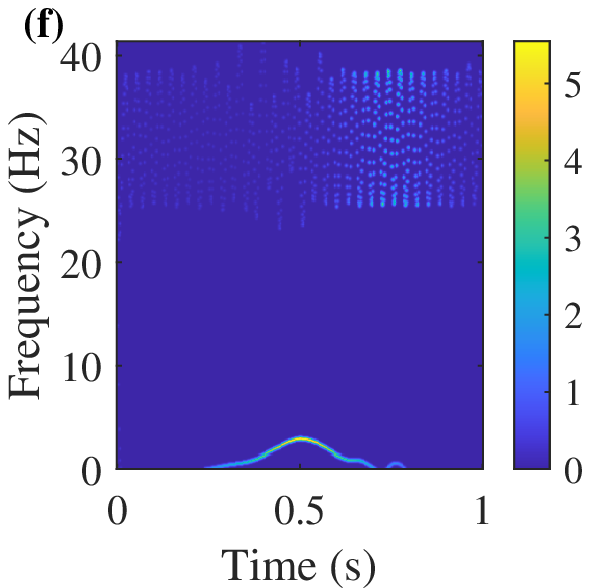}\hfill{}\subfloat{\centering{}\includegraphics[scale=0.6]{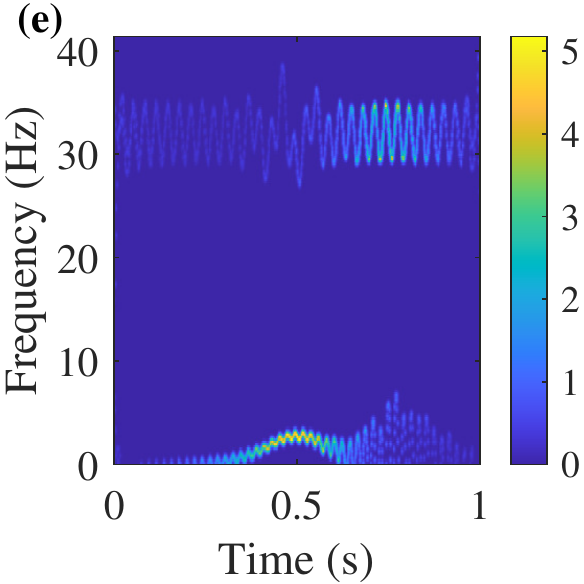}}
\par\end{centering}
\caption{\label{fig:15}TFR results of $f_{\text{Sig3}}(t)$ by: (a) EFD, (b)
EWT, (c) FDM-LTH, (d) FDM-HTL, (e) VMD, (f) EMD, and (g) benchmark.}
\end{figure}

TFRs of $f_{\text{Sig3}}(t)$ of components decomposed by the six
methods are compared to further study their performances. TFR of decomposed
components by the EFD, EWT, VMD and EMD are calculated using Hilbert
transform and those by the FDM-HTL and FDM-LTH are directly obtained
as instantaneous amplitude and frequency. A benchmark TFR is obtained
by using Hilbert transforms of theoretical components. TFRs corresponding
to six decomposition methods and the benchmark one are shown in Fig.
\ref{fig:15} and RMSEs corresponding to TFR are calculated and listed
in Table \ref{tab:5}. From Fig. \ref{fig:15}, it can be found that
the TFRs associated with the EFD and FDM-LTH compare well with the
benchmark one, while the comparisons between the TFRs associated with
the EWT and EMD and the benchmark one are acceptable. However, large
differences can be observed for the TFRs corresponding to the VMD
and FDM-HTL with the benchmark one.

In addition, RMSEs of between magnitudes in the TFRs by the six methods
and those in the benchmark TFR at all frequencies and times are calculated
and listed in Table \ref{tab:5}. It can be seen that the RMSEs associated
with the TFRs by the EFD and EWT are the smallest. Those associated
with the TFR by the EMD are relatively small, while those associated
with the TFRs by the FDM-LTH, FDM-HTL and VMD are large. The TFRs
shown in Fig. \ref{fig:15} and the RMSEs in Table \ref{tab:5} show
that the EFD can yield accurate TFR, and the inconsistency between
TFR results by the FDM-LTH and FDM-HTL is observed. 

\begin{table}
\caption{\label{tab:5}Calculated RMSEs of TFR associated with $f_{\text{Sig3}}(t)$.}

\centering{}{\footnotesize{}}%
\begin{tabular}{cccccc}
\hline 
{\footnotesize{}EFD} & {\footnotesize{}EWT} & {\footnotesize{}FDM-LTH} & {\footnotesize{}FDM-HTL} & {\footnotesize{}VMD} & {\footnotesize{}EMD}\tabularnewline
\hline 
{\footnotesize{}$1.24\times10^{-1}$ } & {\footnotesize{}$1.24\times10^{-1}$ } & {\footnotesize{}$1.35\times10^{-1}$ } & {\footnotesize{}$1.38\times10^{-1}$ } & {\footnotesize{}$1.36\times10^{-1}$ } & {\footnotesize{}$1.29\times10^{-1}$ }\tabularnewline
\hline 
\end{tabular}{\footnotesize\par}
\end{table}

\subsection{Computational cost}

To explore the computational cost of the EFD, computation times by
the EFD, EWT, FDM, VMD and EMD for $f_{\text{Sig3}}(t)$, $f_{\text{Sig4}}(t)$
and $f_{\text{Sig5}}(t)$ are listed in Table \ref{tab:6}. All computations
are conducted on MATLAB R2020a on a PC with an Intel Xeon W-2123 CPU,
16.0 GB of RAM and 64-bit Windows 10. It can be seen that the EFD
requires the shortest computation time among the six methods. The
computation times associated with the EFD and EWT are comparable and
while those of the FDM-HTL and FDM-LTW are large. The computational
times of the VMD and EMD depended on their parameters; though they
can greatly vary, they are longer than that of the EFD. Hence, it
can be concluded that the EFD is the most computationally efficient. 

\begin{table}
\caption{\label{tab:6}Computation time for $f_{\text{Sig3}}(t)$, $f_{\text{Sig4}}(t)$
and $f_{\text{Sig5}}(t)$ by the EFD, EWT, FDM-LTH, FDM-HTL, VMD and
EMD.}

\centering{}{\footnotesize{}}%
\begin{tabular}{ccccccc}
\hline 
\multirow{2}{*}{{\footnotesize{}Signal}} & \multicolumn{6}{c}{{\footnotesize{}Computation Time (s)}}\tabularnewline
\cline{2-7} \cline{3-7} \cline{4-7} \cline{5-7} \cline{6-7} \cline{7-7} 
 & {\footnotesize{}EFD} & {\footnotesize{}EWT} & {\footnotesize{}FDM-LTH} & {\footnotesize{}FDM-HTL} & {\footnotesize{}VMD} & {\footnotesize{}EMD}\tabularnewline
\hline 
{\footnotesize{}$f_{\text{Sig3}}(t)$} & {\footnotesize{}$1.66\times10^{-2}$} & {\footnotesize{}$6.33\times10^{-2}$} & {\footnotesize{}$1.75\times10^{-1}$} & {\footnotesize{}$1.09\times10^{-1}$} & {\footnotesize{}$1.75\times10^{-1}$} & {\footnotesize{}$8.21\times10^{-2}$ }\tabularnewline
{\footnotesize{}$f_{\text{Sig4}}(t)$} & {\footnotesize{}$1.54\times10^{-2}$} & {\footnotesize{}$6.61\times10^{-2}$} & {\footnotesize{}$2.59\times10^{-1}$} & {\footnotesize{}$1.04\times10^{-1}$} & {\footnotesize{}$1.96\times10^{-0}$} & {\footnotesize{}$9.58\times10^{-2}$ }\tabularnewline
{\footnotesize{}$f_{\text{Sig5}}(t)$} & {\footnotesize{}$1.67\times10^{-2}$ } & {\footnotesize{}$7.02\times10^{-2}$ } & {\footnotesize{}$1.59\times10^{-1}$ } & {\footnotesize{}$1.16\times10^{-1}$ } & {\footnotesize{}$1.41\times10^{-0}$ } & {\footnotesize{}$1.51\times10^{0}$ }\tabularnewline
\hline 
\end{tabular}{\footnotesize\par}
\end{table}

\section{Concluding remarks}

In this paper, an accurate and efficient EFD method is proposed to
decompose time-domain signals. The EFD consists of two critical steps:
the improved segmentation technique and construction of a filter bank.
In the improved segmentation technique, only meaningful Fourier spectrum
components are included and the last segment that can contain noise
is narrowed, which can decrease the size of the segment and alleviate
adverse effects of noise. In a constructed filter bank, transition
phases of filter functions are eliminated, which can improve the decomposition
performance for closely-spaced modes. Two numerical investigations
are conducted on non-stationary signals, it is shown that the EFD
can yield decomposition results with high accuracy and consistency.
Two numerical investigations are conducted on two stationary signals
to study the decomposition performance of the EFD for closely-spaced
modes. It is shown that the EFD can yield decomposition results for
the closely-spaced modes with high accuracy and consistency and its
decomposition results are more accurate than those by the other decomposition
methods. In addition, it is shown that the EFD can yield accurate
TFRs for non-stationary signals. Comparisons between computation times
by the EFD, EWT, FDM, VMD and EMD show that the EFD is the most computationally
efficient. A future work can be an investigation of the applicability
of the EFD to signals/data of higher dimensions, such as digital images. 

A MATLAB implementation of the proposed algorithm will be available
at the MATLAB Central. 

\section*{Acknowledgments}

The author Y.F. Xu is grateful for the financial support from the
National Science Foundation through Grant No. CMMI-1762917. The authors
gratefully acknowledge valuable discussions with and input from Dr.
Gang Yu, Dr. Pushpendra Singh, Dr. Shiqian Chen and Dr. Heng Li. They
also thank people who share their contributions to the signal processing
community. 

\section*{Conflicts of Interest }

The authors declare no conflict of interest.

\bibliographystyle{elsarticle-num}
\bibliography{reference}

\end{document}